\documentclass[reprint, prd]{revtex4-2}
\usepackage{amsmath, amssymb}
\usepackage{graphicx,color,float,tabularx,siunitx}
\usepackage[caption=false]{subfig}
\usepackage{placeins}
\usepackage{rviewport}
\usepackage{orcidlink}
\definecolor{colorLink}{rgb}{0,0,180} 
\usepackage{hyperref}
\hypersetup{
   colorlinks = true,
   citecolor  = colorLink,
   urlcolor   = colorLink,
   linkcolor  = colorLink,
}

\DeclareSIUnit\year{yr}

\usepackage[sort&compress]{natbib}

\usepackage[shortcuts]{extdash}

\begin{document}
\title{Holographic Fractional Order Phase Transitions in CFTs Dual to AdS Black Holes}

\author{Abhishek~Baruah$^{1,2}$~\orcidlink{0009-0006-2069-0872}}

\email{rs\_abhishekbaruah@dibru.ac.in}
\author{Prabwal~Phukon$^{1,3}$~\orcidlink{0000-0002-4465-7974}}
\email{prabwal@dibru.ac.in}
\affiliation{$^1$Department of Physics, Dibrugarh University, Dibrugarh, 786004, 
Assam, India}
   \affiliation{$^2$Department of Physics, Patkai Christian College, Ch$\ddot{u}$moukedima, 797103, Nagaland, India.}
   \affiliation{$^3$Theoretical Physics Division, Centre for Atmospheric Studies, Dibrugarh University, Dibrugarh, 786004, Assam, India}
\begin{abstract}
In this work, we investigate the CFT phase transitions of various AdS black hole solutions including the Reissner–Nordström–AdS (RN-AdS) black hole, the ModMax-AdS black hole, and the RN-AdS black hole formulated within the framework of Kaniadakis statistics through the lens of the AdS/CFT correspondence. Employing the generalized Ehrenfest classification scheme based on fractional-order derivatives, we analyze the nature of phase transitions at both Davies points and critical points. Davies points, defined as the loci of divergent heat capacity, are typically associated with second-order transitions in the classical Ehrenfest paradigm. However, a refined analysis reveals that these points can be categorized into two distinct types: the first corresponds to extrema in the temperature profile, while the second aligns with its inflection point, i.e., the thermodynamic critical point. Our findings demonstrate that the order of the phase transition is sensitive to this classification, with the first type corresponding to a fractional order of 3/2, and the second to 4/3, which is the same for the RN-AdS black holes. Notably, when a specific constraint is imposed, we observe a 3/2-order phase transition for both the RN-AdS and ModMax-AdS black holes, whereas in the case of the RN-AdS black hole with Kaniadakis statistics, two critical points arise under constrained paths, each exhibiting a transition of order 4/3. This generalized, fractional-order framework enables a more precise and discriminating characterization of CFT phase transitions in holographic settings, revealing distinctions that remain hidden under traditional classifications. The results provide deeper insight into the rich structure of black hole thermodynamics on the CFT side and highlight the significance of fractional calculus as a powerful tool for probing critical phenomena within the AdS/CFT framework.
\end{abstract}


\maketitle                                                                      

\section{Introduction}

The pioneering contributions of Hawking and Bekenstein established a robust framework for exploring the thermodynamic properties of black holes using the formula 
\begin{equation}
T=\frac{\kappa}{2\pi},\quad S=\frac{A}{4G_N}
\end{equation}
Here, 
$\kappa$, $A$, and $G_N$
  denote the surface gravity, the area of the event horizon, and Newton's universal gravitational constant, respectively \cite{hawking_rad,hawking_explosions,hawking_particle_creation,bardeen_carter_hawking,bekenstein_entropy,wald_entropy,bekenstein_pt,wald_lrr}
. Given that phase transitions and critical phenomena are fundamental aspects of conventional thermodynamic systems, it is natural to investigate analogous behaviors in black hole physics. In this context, Davies demonstrated that the heat capacity of certain black hole solutions exhibits divergences at specific points, which, in accordance with the Ehrenfest classification, are indicative of second-order phase transitions \cite{hut_charged,davies_thermo,sokolowski_mazur}. These divergence points are now widely referred to as Davies points.\\
In general, the heat capacity of a black hole with respect to a thermodynamic variable $X$ can be expressed as a function of the horizon radius 
$r_h$ , temperature $T$, and entropy $S$, typically in the form
\begin{equation}
\mathcal{C}_X=T\left(\frac{\partial S}{\partial T}\right)_X=\frac{T\left(\frac{dS}{d r_h}\right)}{\left(\frac{dT}{d r_h}\right)}
\end{equation}
Here, $X$ denotes a specific thermodynamic quantity held fixed. A divergence in the heat capacity occurs when the derivative of temperature with respect to the horizon radius vanishes, i.e., 
$\frac{dT}{dr_h}=0$. This condition corresponds to a thermodynamic instability and often signals a phase transition. In particular, when the temperature curve exhibits an inflection point with respect to 
$r_h$, it marks the location of a Davies point, which coincides with a second-order critical point in black hole thermodynamics.\\
The behavior of various black hole solutions in the vicinity of critical points has been the subject of extensive investigation. Within the framework of the AdS/CFT correspondence, the phase structure of charged AdS black holes has been explored in detail \cite{PhysRevD.60.064018,PhysRevD.60.104026}. Additionally, the thermodynamic stability and phase transitions of different charged black holes have been analyzed by enclosing them within a finite boundary or cavity, thereby simulating equilibrium conditions \cite{peca_lemos,carlip_vaidya,lu_roy_xiao}. An alternative approach employs thermodynamic geometry, wherein the thermodynamic curvature serves as an indicator of phase transitions near criticality \cite{liu_lu_luo_shao,banerjee_modak_samanta,cao_chen_shao_prd83,wang_chen_jing,niu_tian_wu,wei_liu_prd87}. Complementing this, the framework of thermodynamic topology has emerged as a powerful method to characterize phase structures through topological invariants such as the Euler characteristic, offering a global perspective on critical behavior and transitions \cite{wei_liu_topology,wei_liu_mann_topology,hazarika_phukon_ptep,gogoi_phukon_eh_ads,hazarika_phukon_hl_ads,hazarika_gogoi_phukon_hawking_page}. These geometric and topological techniques collectively enrich our understanding of black hole phase transitions beyond standard thermodynamic descriptions. Motivated by the perspective introduced in \cite{dolan_2011} known as the Extended Phase Space Thermodynamics, which interprets the cosmological constant as a pressure term and its conjugate as a thermodynamic volume, the concept has been developed. This framework enables a direct analogy with classical liquid–gas systems, allowing for the examination of 
$P-V$ criticality and the associated phase structure in a wide class of AdS black holes \cite{kubiznak_mann,cai_cao_li_2013,hendi_vahidinia,chen_liu_liu,mo_zeng_li_2013,liu_zou_wang_2014,ma_liu_zhao_2014,xu_xu_zhao_2014,ma_ma_2015,xu_cao_hu_2015,wei_cheng_liu_2016,ma_wang_2017,xu_2021}. However, despite its successes, the EPST framework is not without challenges, particularly the ensemble ambiguity and the lack of homogeneity in the energy function, both of which obscure the physical interpretation of black hole thermodynamic variables. To address these limitations, the Restricted Phase Space Thermodynamics (RPST) was proposed \cite{gao_zhao_rpst}, wherein the cosmological constant is held fixed and the first law is reformulated in terms of boundary CFT quantities such as the central charge 
$C$, CFT volume 
$\mathcal{V}$, and their conjugates. This approach restores homogeneity, resolves ensemble inconsistencies, and allows for a consistent CFT based interpretation of phase transitions in AdS black holes \cite{gao_kong_zhao_kerr,sadeghi_rps_taubnut_2022,ali_ghosh_wang_kerrsen,awal_phukon_ned_rps,ladghami_rps_egb,sadeghi_rps_taubnut_cc,sadeghi_rps_topology,alipour_sadeghi_flux_rpst,baruah_phukon_kaniadakis,baruah_phukon_entropy_models}. Across these studies, it has been consistently concluded that the phase transition occurring at the critical point corresponds to a second-order transition.\\
The AdS/CFT correspondence, originally proposed by Maldacena \cite{maldacena_ads_cft}, establishes a powerful duality between gravitational theories in asymptotically Anti–de Sitter (AdS) spacetimes and conformal field theories (CFTs) defined on their boundaries. This correspondence has transformed black hole thermodynamics by offering a quantum field theoretic interpretation of gravitational phenomena. Specifically, black hole observables such as mass 
$M$, charge $Q$, temperature 
$T$, and entropy 
$S$ find dual descriptions in the CFT as energy 
$E$, electric potential 
$\Phi$, and central charge 
$C$ \cite{maldacena_ads_cft,visser_2022,karch_2015,mancilla_2024,cong_kubiznak_mann_2021}.\\
Building on this framework, the thermodynamic first law for charged AdS black holes can be reformulated in the CFT language as
\begin{equation}
dE=\mathcal{T}d\mathcal{S}+\varphi d\mathcal{Q}-p d\mathcal{V}+\mu C
\end{equation}
with the Smarr relation expressed as
\begin{equation}
E=\mathcal{T}\mathcal{S}+\varphi \mathcal{Q}+\mu C
\end{equation}
where 
$\mu$ is the chemical potential conjugate to the central charge 
$C$, and 
$p$, 
$\mathcal{V}$ represent the CFT pressure and volume, respectively. These variables are interpreted holographically via a conformal factor 
$\omega$ as
\begin{equation}
E=\frac{M}{\omega},\quad \mathcal{T}=\frac{T}{\omega},\quad \varphi=\frac{\Phi \sqrt{G}}{\omega l},\quad \mathcal{Q}=\frac{Q l}{\sqrt{G}}
\label{eq:dictionary}
\end{equation}
with 
$\omega=R/l$ defining the CFT boundary geometry \cite{visser_2022,karch_2015,mancilla_2024,cong_kubiznak_mann_2021,cong_kubiznak_mann_jhep2022,ahmed_cong_kubiznak_jhep2023,baruah_phukon_prd2025,baruah_phukon_dyonic_ensemble,yang_2025}.\\
Hilfer introduced a generalized version of the Ehrenfest classification scheme, which categorizes phase transitions based on the continuity properties of fractional-order derivatives of the free energy \cite{hilfer1, hilfer2}. Extending this framework, we investigate the CFT phase transition behavior of various black hole systems, including the Reissner–Nordstr\"om AdS (RN-AdS) black hole, the ModMax-AdS black hole, and the RN-AdS black hole modified by Kaniadakis statistics using the AdS/CFT dictionary. \\
Nonlinear electrodynamics (NLED) extends Maxwell’s classical theory to regimes of strong electromagnetic fields, where linear approximations break down. It was originally motivated by the need to eliminate the singularities at point charge locations and to incorporate quantum corrections to classical electrodynamics. This development began with the effective Lagrangian introduced by Heisenberg and Euler \cite{heisenberg_euler_1936}, extended by Schwinger through QED \cite{schwinger_1951}, and refined by Adler in the context of photon–photon interactions \cite{adler_1971}. The coupling of NLED with general relativity has been proposed to model cosmic inflation and construct regular black hole solutions. One of the most prominent NLED models is Born–Infeld (BI) theory, inspired by string theory, which regularizes the self-energy of point charges but does not respect conformal invariance and electromagnetic duality. To address this, the ModMax theory was proposed by Bandos \textit{et al.} \cite{bandos_modmax_2020}, offering a duality-invariant and conformally invariant generalization of Maxwell’s theory. Various black hole solutions have since been derived from this model, including Taub–NUT \cite{bordo_modmax_taubnut}, accelerating \cite{barrientos_modmax_acc}, and dyonic black holes \cite{pantig_modmax_dyonic}, as well as those in modified gravity scenarios \cite{babaei_modmax_frgravity}.\\
 Despite the remarkable success of black hole thermodynamics, the non-extensive nature of Bekenstein–Hawking entropy continues to pose a conceptual challenge. Unlike classical thermodynamic entropy, which is both additive and extensive, black hole entropy violates these foundational principles, thereby motivating the exploration of alternative entropy frameworks \cite{nojiri_odintsov_faraoni1, nojiri_odintsov_paul1, nojiri_odintsov_paul2, nojiri_odintsov2,nojiri_2021,elizalde_2025, kumar_entropy_rel, bialas_czyz_rnyi, huang_zhou_jhep, brustein_medved, nishioka_jhep2014, czinner_iguchi_b, dong_ncomms12472, wen_ijmpd2017, czinner_iguchi_eurpjc2017, qolibikloo_ghodsi, johnson_ijmpd2019, tannukij_wongjun_hirunsirisawat_deesuwan_promsiri, promsiri_hirunsirisawat_liewrian_2020, samart_channuie_2020, ren_jhep_may2021, mejrhit_hajji_2020, nakarachinda_tannukij_et_al_2021, abreu_ananiasneto_epl2021}. In this regard, non-additive entropy formulations such as Rényi and Tsallis entropies have offered valuable perspectives for describing systems with inherent non-extensive characteristics \cite{renyi1960, tsallis1988}
. More recent advancements have introduced further generalizations, including the Barrow entropy \cite{barrow2020}
, Sharma–Mittal entropy \cite{sayahianjahromi2018}
, and the Kaniadakis entropy \cite{kaniadakis2005,drepanou2021}, each proposing modifications aimed at overcoming the limitations of standard entropy definitions. In particular, the Kaniadakis entropy is defined as
\begin{equation}
S=\frac{1}{\kappa}\sinh\left(\kappa \frac{\pi r^2}{G_N}\right)
\end{equation}
where $\kappa$ is a deformation parameter. In the limit 
$\kappa \rightarrow 0$, the Kaniadakis entropy smoothly reduces to the familiar Bekenstein–Hawking entropy, thereby ensuring consistency with the classical formulation.\\
By applying the generalized Ehrenfest approach, it is found that the phase transition at the critical point exhibits a fractional order, differing from the conventional integer-order classification \cite{mengsen_ma}. Notably, this method is primarily utilized to characterize transitions occurring at critical (Davies) points and does not alter the established phase structure for the CFT of black holes. Chabab and Iraoui explored AdS black holes in higher-dimensional spacetimes and those lacking spherical symmetry. Their findings reveal that the order of fractional-phase transitions can indeed be sensitive to the underlying spacetime geometry \cite{chabab_iraoui1}. In a related study, Wang, He, and Ma investigated the fractional phase transitions of Reissner–Nordström–AdS (RN-AdS) black holes at their Davies points \cite{wang_he_ma_fractional_rnads}
. While several studies have analyzed these systems from the gravitational perspective, no comprehensive analysis has yet been performed from the dual CFT side. This gap motivates a deeper exploration of their holographic phase structure within the AdS/CFT correspondence.\\
It is important to note that both categories of Davies points are characterized by divergent heat capacities. As a result, conventional thermodynamic analysis based solely on heat capacity fails to distinguish between them. To address this limitation, we adopt a generalized version of the Ehrenfest classification based on fractional-order derivatives of the free energy. This extended framework offers a more refined toolset for probing the intricate phase structure of thermodynamic systems.\\
As a case study, we investigate the CFT of the three black hole systems to examine the phase structure near its Davies points using the AdS/CFT correspondence. Since the temperature profile of the RN-AdS black hole exhibits qualitatively different behavior at the two types of Davies points in black holes, it is reasonable to anticipate corresponding differences in the underlying phase transitions in the CFT. We therefore ask whether the generalized Ehrenfest classification based on fractional derivatives can effectively distinguish between these two types of Davies points in the CFT side\\
Furthermore, we seek to determine whether the order of the phase transition at the critical point remains consistent when analyzed using the AdS/CFT correspondence from the EPST framework for the three systems. This investigation aims to provide deeper insight into the nature of black hole phase transitions and the role played by generalized thermodynamic frameworks. The paper is organized as follows- We first briefly review the RN-AdS black hole and describe several thermodynamic quantities along with its CFT counterpart in Section \ref{sec:RNADS}. In Section \ref{sec:RNADSF1}, we analyze the phase structures at the first type of Davies point, referring to two Davies points on the left-hand side and right-hand side. In Section \ref{sec:RNADSF2}, we discuss the phase structures of the RN-AdS black hole at the critical point. In Section \ref{sec:MOD} we review the ModMax AdS black hole and describe its thermodynamics of its CFT counterpart. Section \ref{sec:MODF1} and \ref{sec:MODF2} refers to the first and second types of Davies point. In Section \ref{sec:KAN} we discuss the phase transition of RN-AdS using the Kaniadakis entropy. Section \ref{sec:KANF1} and \ref{sec:KANF2} describes the first and second types of Davies point. Finally, in Section \ref{sec:conclusion} we summarize our results and discuss possible future studies.
\section{RN ADS BLACK HOLE}\label{sec:RNADS}
\begin{figure}[htbp]
  \centering
  \subfloat[$\mathcal{C}-\mathcal{S}$ plots]{%
    \includegraphics[width=0.3\textwidth]{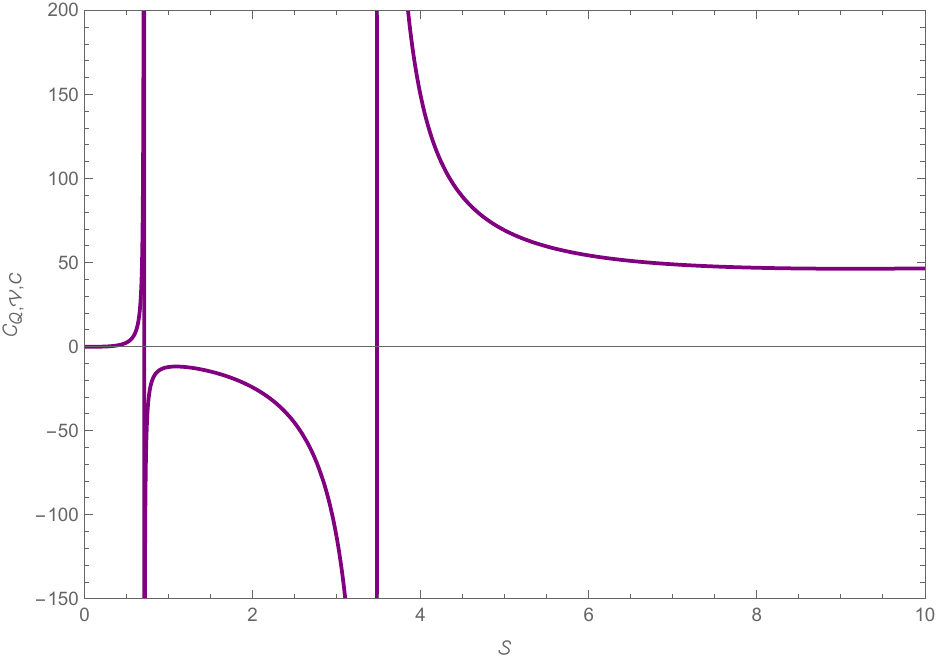}%
    \label{fig:fig1}
  }%
  \hfill
  \subfloat[$\mathcal{C}-\mathcal{S}$ plots]{%
    \includegraphics[width=0.3\textwidth]{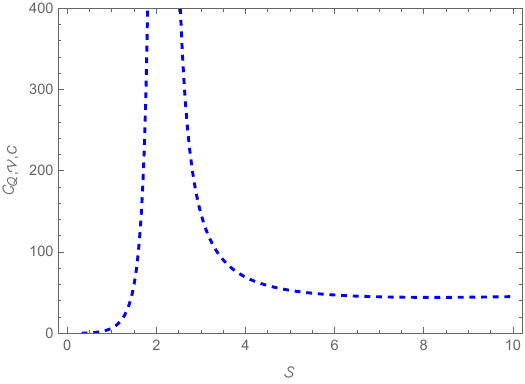}%
    \label{fig:fig2}
  }
  \caption{The heat capacity of the RN-AdS black hole is plotted as a function of the horizon radius \( r_h \) in the CFT thermodynamic framework. The parameters are chosen as \( C = 1 \). The purple and blue dashed curves represent the cases for \( \mathcal{Q} = 1/2 \) and \( \mathcal{Q} = 2/3 \), where \( \mathcal{Q} \) denotes the rescaled CFT charge.
}
  \label{fig:F1}
\end{figure}

We begin by briefly revisiting the class of spherically symmetric, charged black hole solutions in asymptotically anti-de Sitter (AdS) spacetime. These solutions arise within the framework of Einstein–Maxwell theory incorporating a negative cosmological constant $\lambda$. The corresponding action in $D=4$ spacetime dimensions is given by
\begin{equation}
I=\frac{1}{16\pi G_N}\int d^4x\sqrt{-g}(R-2\Lambda-\mathcal{F}^2)
\end{equation}
where $\Lambda$ denotes the cosmological constant, $G_N$ represents the Newtonian gravitational constant, $\mathcal{F}$ is the field strength tensor associated with the $U(1)$ gauge field, and $R$ corresponds to the Ricci scalar curvature. The line element describing a Reissner–Nordstr\"om AdS black hole in static coordinates is given by the following metric
\begin{equation}
ds^2=-f(r)dt^2+\frac{dr^2}{f(r)}+r^2 d\Omega^2_2
\end{equation}
The function $f(r)$ is given by
\begin{equation}
f(r)=1-\frac{2 GM }{r}+\frac{GQ^2}{r^2}+\frac{r^2}{l^2}
\end{equation}
where $M$ is the ADM mass, $Q$ is the electric charge, $r$ is the horizon radius and $l$ is the AdS length. By imposing the condition $f(r)=0$, one can determine the mass $M$ as a function of the horizon radius as
\begin{equation}
M=\frac{G l^2 Q^2+l^2 r^2+r^4}{2 G l^2 r}
\end{equation}
The Hawking temperature and entropy, which is connected to the surface gravity through the relation $T=\kappa/4\pi$ and $S=A/4G$ can thus be written as
\begin{equation}
T=\frac{-G l^2 Q^2+l^2 r^2+3 r^4}{4 \pi  l^2 r^3},\quad S=\frac{\pi r^2}{G}
\end{equation}
To incorporate the boundary central charge into the thermodynamic first law, we consider the holographic duality relation that links the AdS length scale $l$, Newton’s constant $G$, and the central charge $C$, which yields
\begin{equation}
C=\frac{\Omega_2 l^2}{16 \pi G}
\label{eq:ccharge}
\end{equation}
The mass in the CFT thermodynamic framework for a singly charged, rotating black hole is considered. The corresponding spacetime metric is defined as
\begin{equation}
ds^2=\omega^2(-dt^2+l^2d\Omega^2_2)
\end{equation}
The line element of the two-sphere is denoted by $d\Omega^2_2$, and $\omega$ represents an arbitrary, dimensionless conformal factor. In certain studies, this conformal factor is chosen as 
$\omega=R/l$ , where $R$ corresponds to the curvature radius. The parameter $\omega$ facilitates the analysis of a holographic first law wherein Newton’s constant remains fixed while the cosmological constant 
$\Lambda$ is allowed to vary. The spatial volume of the boundary sphere is then expressed as 
\begin{equation}
\mathcal{V}=\Omega_2(\omega l)^2
\label{eq:volume}
\end{equation}
For the volume $\mathcal{V}$ defined within the CFT framework, a corresponding pressure $p$ can be introduced, leading to the presence of a work term $-pd\mathcal{V}$. Utilizing the holographic dictionary, the bulk quantities such as mass, entropy, temperature, electric potential, and charge along with their respective dual counterparts are identified as follows
\begin{equation}
E=\frac{M}{\omega},\quad \mathcal{T}=\frac{T}{\omega},\quad \mathcal{S}=S,\quad \varphi=\frac{\Phi \sqrt{G}}{\omega l},\quad \mathcal{Q}=\frac{Q l}{\sqrt{G}}
\label{eq:holo}
\end{equation}
By using the above equation, the CFT mass is given as
\begin{equation}
E=\frac{4 \pi C \mathcal{S}+\pi ^2 \mathcal{Q}^2+\mathcal{S}^2}{2 \pi   \sqrt{C\mathcal{S V}}}
\end{equation}
The thermodynamic first law is connected to the Smarr relation through the following expression
\begin{equation}
dE=\mathcal{T}dS+\varphi  d\mathcal{Q}-pd\mathcal{V}+\mu dC, \quad E=\mathcal{T}\mathcal{S}+\varphi\mathcal{Q}+\mu C
\end{equation}
The CFT temperature is calculated as
\begin{equation}
\mathcal{T}=\left(\frac{\partial E}{\partial \mathcal{S}}\right)=\frac{ \left(4 \pi  c S-\pi ^2 \mathcal{Q}^2+3 S^2\right)}{4 \pi  \sqrt{C\mathcal{V}S^3}}
\label{eq:temp}
\end{equation}
The heat capacity at fixed charge is defined by the following expression
\begin{equation}
\mathcal{C}_{\mathcal{Q}}=\frac{\partial E}{\partial \mathcal{T}}=\mathcal{T}\frac{\partial \mathcal{S}}{\partial \mathcal{T}}=\frac{2 S \left(4 \pi  C S-\pi ^2 \mathcal{Q}^2+3 S^2\right)}{3 \left(\pi ^2 \mathcal{Q}^2+S^2\right)-4 \pi  C S}
\end{equation}
By analyzing the denominator, the locations of the Davies points can be determined as follows.
\begin{equation}
\mathcal{S}=\frac{2 \pi  C}{3}\pm\frac{1}{3} \sqrt{4 \pi ^2 C^2-9 \pi ^2 \mathcal{Q}^2}
\end{equation}
When $\mathcal{Q}<\frac{2 \sqrt{C^2}}{3}$, two distinct roots emerge, corresponding to the first type of Davies point. In contrast, when $\mathcal{Q}=\frac{2 \sqrt{C^2}}{3}$, these roots merge, signifying the critical point, which is classified as the second type of Davies point.\\
As illustrated in Figure \ref{fig:F1}, two Davies points of the first type are observed along the purple curve. Each of these points demarcates the temperature into two distinct branches, across which the heat capacity exhibits opposite signs. When the two Davies points converge, as shown on the blue-dashed curve, they coalesce into a critical point. The corresponding behavior of the heat capacity is also presented in this figure, revealing that it diverges at both types of Davies points. According to the conventional Ehrenfest classification, these transitions are identified as second-order. In the subsequent sections, we aim to differentiate these phase structures using our generalized form of the Ehrenfest classification.

\subsection{PHASE TRANSITION ASSOCIATED WITH TH FIRST TYPE OF DAVIES POINT}\label{sec:RNADSF1}
\begin{figure}[htbp]
  \centering
  \subfloat[$D^\alpha F-t$ plots]{%
    \includegraphics[width=0.3\textwidth]{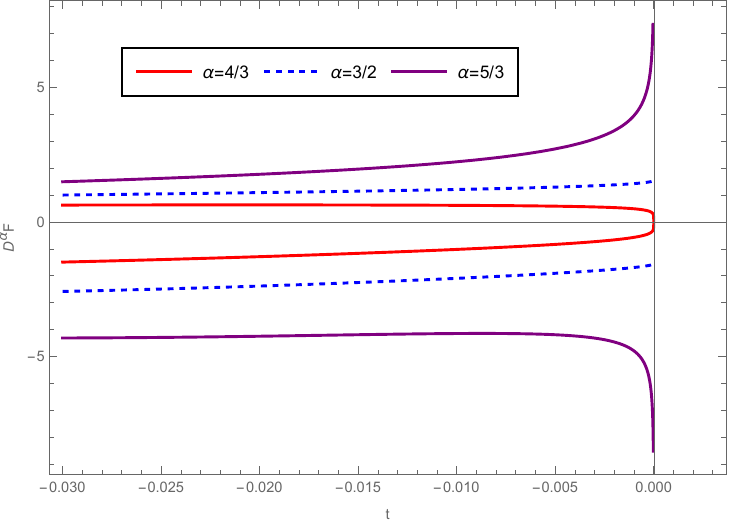}%
    \label{fig:fig1}
  }%
  \hfill
  \subfloat[$D^\alpha F-\rho$ plots]{%
    \includegraphics[width=0.3\textwidth]{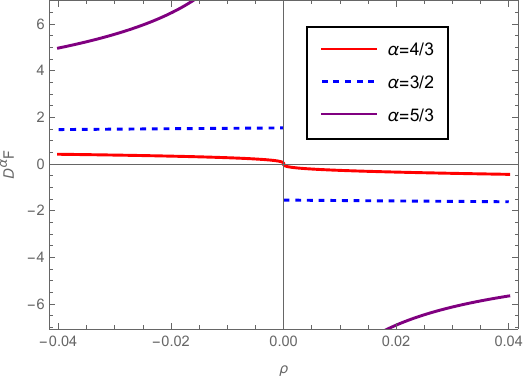}%
    \label{fig:fig2}
  }
  \caption{The behavior of \( D^\alpha_t F(t) \) in the vicinity of the left Davies point is illustrated within the CFT thermodynamic framework for the RN-AdS black hole. The top panel displays the corresponding \( D^\alpha_t F(t) \) vs. \( t \) curves, while the bottom panel presents the \( D^\alpha_t F \) vs. \( \rho \) plots.
}
  \label{fig:F2}
\end{figure}

\begin{figure}[htpb]
  \centering
  \subfloat[$D^\alpha F-t$ plots]{%
    \includegraphics[width=0.3\textwidth]{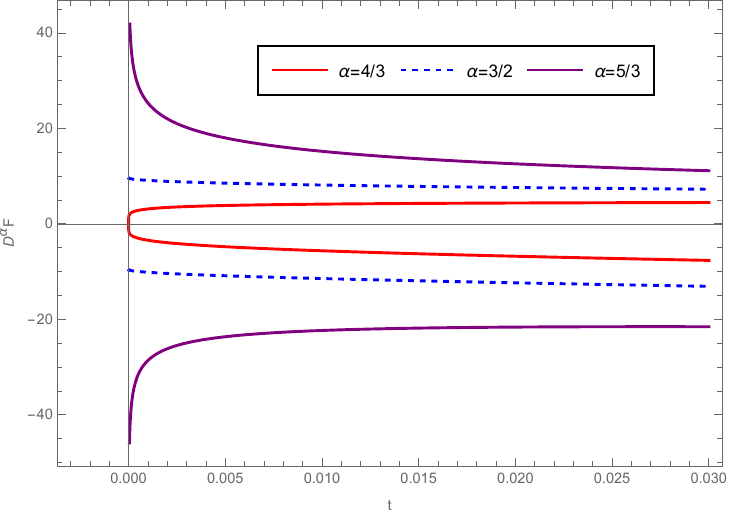}%
    \label{fig:fig1}
  }%
  \hfill
  \subfloat[$D^\alpha F-\rho$ plots]{%
    \includegraphics[width=0.3\textwidth]{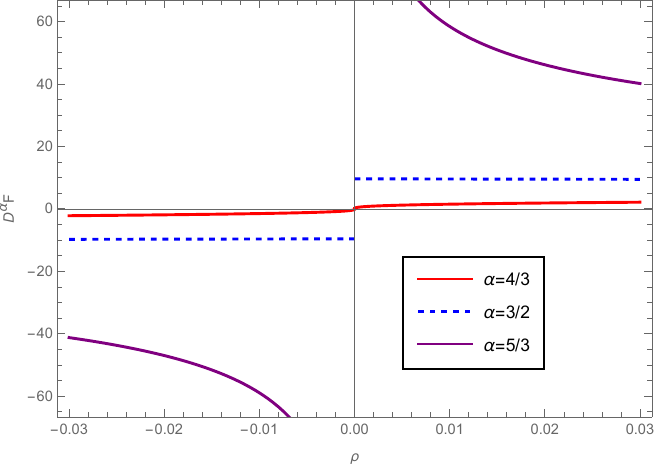}%
    \label{fig:fig2}
  }
  \caption{The behavior of \( D^\alpha_t F(t) \) in the vicinity of the right Davies point is illustrated within the CFT thermodynamic framework for the RN-AdS black hole. The top panel displays the corresponding \( D^\alpha_t F(t) \) vs. \( t \) curves, while the bottom right panel presents the plots of \( D^\alpha_t F \) vs. \( \rho \).
}
  \label{fig:F3}
\end{figure}

As indicated by Eq. \eqref{eq:temp}, for smaller values of $\mathcal{S}$, the temperature of the RN-AdS black hole is primarily influenced by the electric charge $\mathcal{Q}$ and central charge $C$, rather than the CFT volume
$\mathcal{V}$. In contrast, for larger values, the temperature is predominantly governed by 
$\mathcal{V}$. Consequently, the thermal behavior near the left Davies point differs significantly from that near the right Davies point. Hence, it is necessary to examine the phase structures at each Davies point individually.\\
For the sake of simplicity, we fix the parameter value to $\mathcal{V}=C=1$ in this section. The left Davies point is located at
\begin{equation}
\begin{split}
&\mathcal{T}_c=\frac{\sqrt{\frac{3}{\pi }} \left(-3 \mathcal{Q}^2-2 \sqrt{4-9 \mathcal{Q}^2}+4\right)}{\left(2-\sqrt{4-9 \mathcal{Q}^2}\right)^{3/2}}\\
&\mathcal{S}_c=-\frac{1}{3} \pi  \left(\sqrt{4-9 \mathcal{Q}^2}-2\right)
\end{split}
\end{equation}
We introduce the following set of dimensionless variables defined as
\begin{equation}
\mathcal{T}=\mathcal{T}_c(1+t),\quad \mathcal{S}=\mathcal{S}_c(1+\rho)
\label{eq:dim}
\end{equation}
In this scenario, the Davies point is situated at $t=\rho=0$. In this way, a series expansion can be carried out in the vicinity of the Davies point. Since this left Davies point represents the maximum of the temperature profile $t\leq 0$, it consistently corresponds to the highest temperature.\\
Upon setting $\mathcal{Q}=1/2$, a dimensionless form of the equation of state can be derived from  \eqref{eq:temp}
\begin{equation}
\begin{split}
&-\frac{0.822 \rho ^2}{(\rho +1)^{3/2}}-\frac{6.506 \rho }{(\rho +1)^{3/2}}
-\frac{4.337}{(\rho +1)^{3/2}}\\
&+4.337 t+4.337=0
\end{split}
\label{eq:eos}
\end{equation}
In the vicinity of the Davies point, we solve \eqref{eq:eos} to derive a series expansion for $\rho$ expressed in terms of the variable $t$
\begin{equation}
\begin{split}
&\rho=-137.011 t^3+24.119 t^2-10.732 (- t)^{3/2} \\
&-56.442 (- t)^{5/2} -4.958 t  -2.323 \sqrt{-t} \\
&\rho=-137.011 t^3 +24.119 t^2 +10.732 (- t)^{3/2}\\
& +56.442 (- t)^{5/2} -4.958 t  +2.323 \sqrt{- t} 
\end{split}
\label{eq:rho}
\end{equation}
The two series solutions represent the distinct branches on either side of the Davies point. A particularly noteworthy aspect of this result is the presence of terms with fractional powers.\\
The free energy is obtained by using the dimensional quantities \eqref{eq:dim} as
\begin{equation}
\begin{split}
&F=E-\mathcal{T}S\\
&=\frac{\sqrt{\frac{\pi }{3}} \left(9 \left(\rho ^2+2 \rho +4\right) \mathcal{Q}^2+4 \left(\rho ^2-\rho -2\right) \left(\sqrt{4-9 \mathcal{Q}^2}-2\right)\right)}{12 \sqrt{-\left((\rho +1) \left(\sqrt{4-9 \mathcal{Q}^2}-2\right)\right)}}
\end{split}
\end{equation}
By substituting the solutions derived from Eq. \eqref{eq:rho} into the expression for the free energy above, we obtain the following result
\begin{equation}
\begin{split}
&F=1.494-0.751 t -1.163 (- t)^{3/2} +1.8626 t^2 \\
&-3.225 (- t)^{5/2} -6.041 t^3 +\mathcal{O}(t ^{7/2})\\
&F=1.494-0.7514 t +1.163 (- t)^{3/2} +1.862 t^2\\
& +3.225 (-t)^{5/2} -6.041 t^3 +\mathcal{O}(t ^{7/2})
\end{split}
\end{equation}
It is important to highlight the appearance of terms with fractional powers. Notably, the coefficients preceding the term $(-t)^{3/2}$ exhibit opposite signs.\\
Next, we proceed to evaluate the derivatives of the free energy, employing Caputo's formulation of fractional derivatives \cite{caputo_1967,kilbas_marichev_samko_1993,gorenflo_mainardi_1997}. The key expressions required for this analysis are presented below
\begin{equation}
D^\alpha_x x^n=0, \quad D^\alpha_x x^n\propto x^{a-\alpha}
\end{equation}
For further details regarding the properties of fractional derivatives, the reader is referred to the appendix of \cite{mengsen_ma}.\\
The fractional derivative of order $\alpha$ for the free energy, under the condition 
$1<\alpha\leq 2$, takes the following form
\begin{equation}
D^\alpha_t F(t)=(-t)^{1-\alpha}[A(-t)^{1/2}\pm Bt+C(-t)^{3/2} \pm Dt^2+...]
\label{eq:series}
\end{equation}
Here, the symbols $A,B,C,D$ represent constant coefficients. The $``+"$ sign denotes the case where the Davies point is approached from the left, while the $``-"$ sign corresponds to approaching it from the right.\\
Due to the condition $t\leq 0$, the $D^\alpha_t F-t$-curves show up on the left side of the 
$t=0$ axis. Alternatively, 
$D^\alpha_t F-t$ can be expressed as a function of $\rho$, in which case the behavior of $D^\alpha_t F$ may appear on both sides of the $\rho=0$ axis. By inserting specific values of $\alpha$ into  \eqref{eq:series} and examining the limiting behavior as $t\rightarrow 0$, we arrive at the following result
\[
\lim_{t \to 0^-} D_t^\alpha F(t) =
\begin{cases}
0, & \alpha < 3/2, \\
\pm 1.547, & \alpha = 3/2, \\
\pm \infty, & \alpha > 3/2.
\end{cases}
\]
An examination of the behavior of 
$D^\alpha_t F(t)$, as illustrated in Figure \ref{fig:F2}, indicates that for 
$\alpha<3/2$, the function remains continuous at the Davies point; for 
$\alpha=3/2$, it exhibits a finite discontinuity; and for 
$\alpha>3/2$, it diverges. Based on the generalized classification framework, this implies that a phase transition of order 
$\alpha=3/2$ takes place at the Davies point.\\
We now proceed to examine the CFT phase structure of the analogous RN-AdS black hole in the vicinity of the right Davies point. The methodology employed here closely parallels that of the previous analysis. In this scenario, the Davies point is located at
\begin{equation}
\begin{split}
&T_c=\frac{\sqrt{\frac{3}{\pi }} \left(-3 \mathcal{Q}^2+2 \sqrt{4-9 \mathcal{Q}^2}+4\right)}{\left(\sqrt{4-9 \mathcal{Q}^2}+2\right)^{3/2}} \\ &S_c=\frac{1}{3} \pi  \left(\sqrt{4-9 \mathcal{Q}^2}+2\right)
\end{split}
\end{equation}
Utilizing the dimensionless variables from \eqref{eq:dim}, the equation of state simplifies to the following form
\begin{align}
& -\frac{17.991\, \rho^2}{(\rho + 1)^{3/2}} 
  - \frac{57.639\, \rho}{(\rho + 1)^{3/2}} 
  - \frac{38.426}{(\rho + 1)^{3/2}} \nonumber \\
& \quad + 38.426\, t + 38.426 = 0
\label{eq:eos1}
\end{align}
It is worth emphasizing that the right Davies point of the RN-AdS black hole corresponds to the minimum of the temperature profile. Consequently, the condition $t>0$ holds throughout the analysis.\\
Following a similar approach, we solve \eqref{eq:eos1} to express $\rho=\rho(t)$ as a series expansion and subsequently substitute the results into the free energy expression. This yields the free energy in the following form
\begin{align}
F &= 1.664 - 3.309\, t - 7.227\, t^{3/2} - 7.363\, t^2\\
& - 5.636\, t^{5/2} - 2\, t^3 + \mathcal{O}(t^{7/2}) \nonumber \\
F &= 1.664 - 3.309\, t + 7.227\, t^{3/2} - 7.363\, t^2\\
& + 5.636\, t^{5/2} - 2\, t^3 + \mathcal{O}(t^{7/2}) \nonumber
\end{align}
Apart from similarities in the coefficient values and the variable 
$t$, the functional forms of these free energy expressions closely resemble those obtained near the left Davies point. As a result, their corresponding fractional derivatives exhibit similar behavior. However, the explicit forms are omitted here for brevity. In Figure \ref{fig:F3}, we directly present the behavior of the fractional derivatives of the free energy as functions of $t$ and $\rho$. The limiting values of these fractional derivatives are given below
\[
\lim_{t \to 0^+} D_t^\alpha F(t) =
\begin{cases}
0, & \alpha < 3/2, \\
\pm 9.608 , & \alpha = 3/2, \\
\pm \infty, & \alpha > 3/2.
\end{cases}
\]
In conclusion, although the temperature profiles differ at the left and right Davies points, the underlying phase structures at both locations exhibit similar characteristics. Notably, the phase transitions at these points correspond to fractional orders specifically of order $\alpha=3/2$ rather than being first-order, as traditionally presumed. To the best of our knowledge, in conventional thermodynamic systems, temperature typically behaves monotonically and does not exhibit features analogous to those observed at the first type of Davies point.

\subsection{PHASE TRANSITION ASSOCIATED WITH THE SECOND TYPE OF DAVIES POINT}\label{sec:RNADSF2}
We now turn our attention to the second category of Davies point namely, the critical point which is characterized by the condition
\begin{equation}
\frac{d\mathcal{T}}{d\mathcal{S}}=0,\quad \frac{d^2\mathcal{T}}{d\mathcal{S}^2}=0
\end{equation}
The critical points are given as
\begin{equation}
\mathcal{Q}_c= \frac{2 C}{3},\quad \mathcal{S}_c= \frac{2 \pi  C}{3},\quad \mathcal{T}_c=\frac{2 \sqrt{\frac{2}{3 \pi }}}{ \sqrt{\mathcal{V}}}
\end{equation}
We continue to employ the dimensionless variables previously defined in \eqref{eq:dim}. Since the critical point represents a point of inflection, the dimensionless temperature $t$ may take values either greater or less than zero.\\Following the definition provided in \eqref{eq:dim}, we similarly define $\mathcal{Q}=\mathcal{Q}_c(1+q)$. With this substitution, the equation of state simplifies to the following form
\begin{equation}
\label{eq:eos1.1}
q^2+2 q+\frac{3}{16} \rho ^4 (t+1)+\rho ^3 \left(-\frac{t}{2}-\frac{1}{2}\right)+3 \rho ^2 t+12 \rho  t+8 t=0
\end{equation}
This is a quartic equation in the variable $\rho$. Its real solutions are given as follows
\begin{equation}
\begin{split}
&\rho(t)=3.968 t^{2/3} +10.368 t^{4/3} +23.629 t^{5/3} +163.073 t^{7/3}\\
& +452.378 t^{8/3} +1286.31 t^3 +60.554 t^2 +2.519 t^{1/3}+5.75 t
\end{split}
\label{eq:rho1} 
\end{equation}
It is important to emphasize, however, that this series expansion remains valid only within the regime where $2q+8t\neq 0$.
In the vicinity of the critical point, the reduced free energy takes the following form
\begin{equation}
\tilde{F}=F/F_c=\frac{3 q^2-\rho ^2+6 q+4 \rho +8}{8 \sqrt{\rho +1}}
\end{equation}
where $F_c=\frac{4 \sqrt{\frac{2 \pi }{3}} C}{3 \sqrt{\mathcal{V}}}$. Substituting \eqref{eq:rho1} in the reduced free energy we get
\begin{equation}
\begin{split}
&\tilde{F}(q,t)=0.297 q t^{2/3}+(-0.757 q-1.889) t^{4/3}\\
&+(-1.568 q-3.571) t^{5/3}
+(-4.397 q-8.5) t^2\\
&-0.944 q t^{1/3}+(-0.281 q-1) t+\frac{3 q}{4}+1
\end{split}
\label{eq:free}
\end{equation}
Subsequently, we evaluate the fractional derivatives of the free energy and examine their behavior in the limiting case as 
$t\rightarrow 0^+$. The resulting expressions are presented below
\[
\lim_{t \to 0^+} D_t^\alpha \tilde{F}(t) =
\begin{cases}
0, & \alpha < 4/3, \\
 9.00067 , & \alpha =4/3, \\
\pm \infty, & \alpha > 4/3.
\end{cases}
\]
It is observed that the fractional derivative of the free energy of order $4/3$ exhibits a discontinuity, which serves as a hallmark of a phase transition of order $4/3$. We consider $q\rightarrow0$ which is the critical point in \eqref{eq:free} getting 
\[
F(t) =
\begin{cases}
-1.88988 t^{4/3}-3.57165 t^{5/3}-8.5 t^2\\-t+1, & t>0,  \\
7.55953 t^{4/3}+25.5968 t^{5/3}+84.75 t^2\\+ t+1, & t<0.
\end{cases}
\]

\begin{figure}
\includegraphics[scale=0.5]{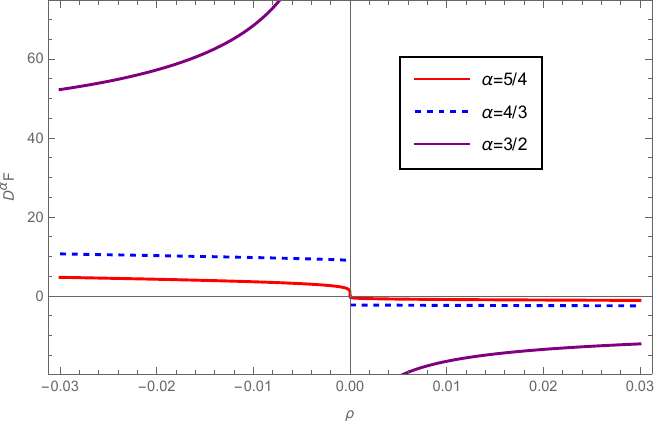}
\caption{ The behavior of \( D^\alpha_t \tilde{F}(\rho) \) near the critical point is shown in the context of CFT thermodynamics for the RN-AdS black hole.
}
\label{fig:F4}
\end{figure}

Due to the presence of the $t^{4/3}$ terms, a finite discontinuity is observed at 
$\alpha=4/3$.
\begin{equation}
\lim_{t\rightarrow 0^-}\frac{d^\alpha \tilde{F}}{dt^\alpha}=-\frac{3.046}{\Gamma \left(\frac{2}{3}\right)}\neq \lim_{t\rightarrow0^+}\frac{d^\alpha \tilde{F}}{dt^\alpha}=\frac{12.188}{\Gamma \left(\frac{2}{3}\right)}
\end{equation}
In accordance with the generalized classification framework, the phase transition occurring at the critical point is identified as being of fractional order $\alpha=4/3$. The corresponding behavior is illustrated in Figure \ref{fig:F4}.\\
Furthermore, it is necessary to address the exceptional case 
$2t+8q=0$, that is in the $(t-q)$ plane, which defines a specific trajectory in the parameter space. The critical point must be approached along this particular direction from both sides. Under this condition, the equation of state \eqref{eq:eos1.1} reduces to the following form
\begin{equation}
\frac{3}{16} \rho ^4 (t+1)+\rho ^3 \left(-\frac{t }{2}-\frac{1}{2}\right)+3 \rho ^2 t  +12 \rho  t+q^2 =0
\end{equation}
The corresponding series expansion solutions are given below
\begin{equation}
\begin{split}
&\rho(t)=-17.0699 t^{3/2}-238.518 t^{5/2}+1002.07 t^3+60 t^2\\
&+7.5 t-4.89898 \sqrt{t}\\
&\rho(t)=17.0699 t^{3/2}+238.518 t^{5/2}+1002.07 t^3+60 t^2\\
&+7.5 t+4.89898 \sqrt{t} 
\end{split}
\end{equation}
In this specific case, the emergence of terms with fractional powers gives rise to a free energy expression that likewise contains fractional exponents. This behavior was not observed in the case of RN-AdS black holes. The reduced free energy at the critical point $q\rightarrow 0$ are
\[
\tilde{F}(t) =
\begin{cases}
7.34847 t^{3/2}+78.8812 t^{5/2}+28.6875 t^3\\+6.75 t^2+1, & t>0,  \\
-7.34847 t^{3/2}-78.8812 t^{5/2}+28.6875 t^3\\+6.75 t^2+1, & t<0.
\end{cases}
\]
We compute the fractional derivatives of the free energy and analyze their behavior in the limiting case as $t\rightarrow 0^+$ and $t\rightarrow 0^-$. The resulting expressions are provided below
\[
\lim_{t \to 0^+} D_t^\alpha F(t) =
\begin{cases}
0, & \alpha < 3/2, \\
 \pm 9.76862 , & \alpha =3/2, \\
\pm \infty, & \alpha > 3/2.
\end{cases}
\]
\begin{figure}
\includegraphics[scale=0.5]{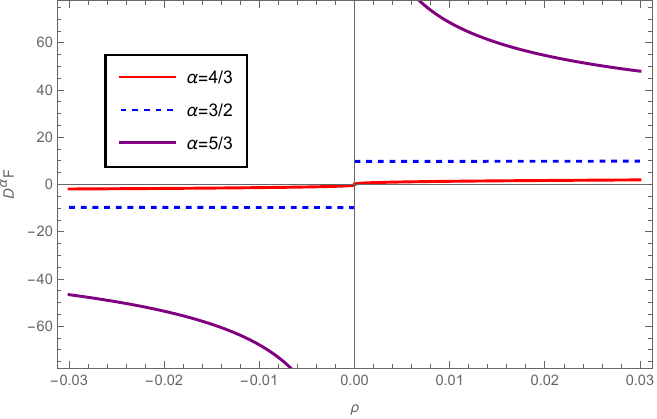}
\caption{The behavior of \( D^\alpha_t \tilde{F}(\rho) \) near the critical point is shown in the context of CFT thermodynamics for the RN-AdS black hole.
}
\label{fig:5}
\end{figure}
As demonstrated above and in Figure \ref{fig:5} the fractional derivative of the free energy of order $3/2$ displays a discontinuity, indicating the occurrence of a phase transition of order 
$3/2$ in the special case where the condition 
$2t+8q=0$ is satisfied which was not seen in its analogous RN AdS black hole.\\
The emergence of different fractional orders $4/3$ without imposing any constraints and 
$3/2$ along the special direction 
$2t+8q=0$ reflects distinct approaches to the critical point in the thermodynamic parameter space. The 
$4/3$ order transition arises generically and represents the dominant critical behavior of the system, making it the physically relevant classification. In contrast, the 
$3/2$ order transition occurs only under a fine-tuned path and should be viewed as a special-case scenario. Therefore, the 
$4/3$ order transition should be considered the primary result.
\section{MOD MAX ADS BLACK HOLES}\label{sec:MOD}
\begin{figure}[htbp]
  \centering
  \subfloat[$\mathcal{C}-\mathcal{S}$ plots]{%
    \includegraphics[width=0.3\textwidth]{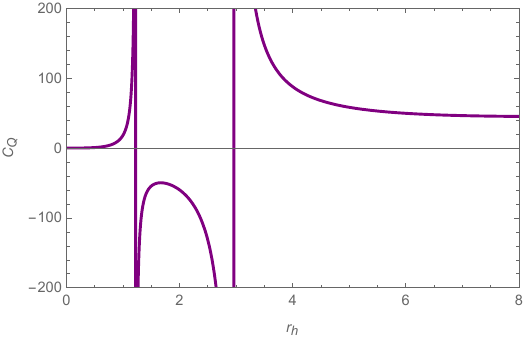}%
    \label{fig:Fig6a}
  }%
  \hfill
  \subfloat[$\mathcal{C}-\mathcal{S}$ plots]{%
    \includegraphics[width=0.3\textwidth]{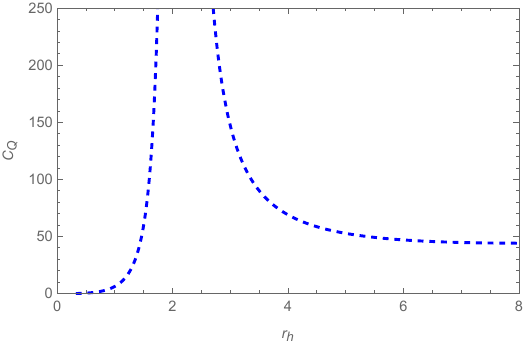}%
    \label{fig:Fig6b}
  }
  \caption{The heat capacity in the CFT thermodynamic framework corresponding to the ModMax-AdS black hole is plotted as a function of the horizon radius. The parameters are chosen as $C=\eta=1$, representing the central charge and the CFT volume, respectively. The blue and red dashed curves correspond to the cases \( \mathcal{Q} = 1 \) and \( \mathcal{Q} = \frac{2 \sqrt{e}}{3} \), respectively.
}
  \label{fig:Fig6}
\end{figure}

The action for Einstein gravity coupled with ModMax electrodynamics and a cosmological constant in four-dimensional spacetime is given by the following expression
\begin{equation}
I=\frac{1}{16 \pi G_N}\int d^4 x\sqrt{-g}(R-2\lambda-4\mathcal{L})
\end{equation}
Here, $R$ denotes the Ricci scalar curvature and $\Lambda$ signifies the cosmological constant. The quantity $g$ refers to the determinant of the metric tensor, i.e., $g=det(g_{\mu \nu})$. In the action presented above, 
$\mathcal{L}$ denotes the Lagrangian density corresponding to Mod Max electrodynamics, and is defined as follows
\begin{equation}
\mathcal{L}=\mathcal{S} \cos \gamma -\sqrt{\mathcal{S}^2+\mathcal{P}^2}\sinh \gamma
\end{equation}
Here, $\gamma$ denotes the ModMax parameter, which is a dimensionless constant. Furthermore, $\mathcal{S}$ and $\mathcal{P}$ correspond to a true scalar and a pseudoscalar quantity, respectively, as defined in the following expressions given as
\begin{equation}
\mathcal{S}=\frac{\mathcal{F}}{4},\quad \mathcal{P}=\frac{\tilde{\mathcal{F}}}{4}
\end{equation}
The quantity 
$\mathcal{F}=F_{\mu \nu} F^{\mu \nu}$ is identified as the Maxwell invariant. Here, 
$F_{\mu \nu}$ denotes the electromagnetic field strength tensor, defined by 
$F_{\mu \nu}=\partial_\mu A_\nu-\partial_\nu A_\mu$, where 
$A_\mu$ is the gauge potential. Additionally, 
$\tilde{\mathcal{F}}=F_{\mu \nu}\tilde{F}^{\mu \nu}$ , with the dual field strength tensor given by $\tilde{F}^{\mu \nu}=\frac{1}{2}\epsilon_{\mu \nu} ^{\rho \lambda}F_{\rho \lambda}$.\\
In this context, we examine a four-dimensional static spacetime described by the following metric ansatz
\begin{equation}
ds^2=-f(r)dt^2+\frac{dr^2}{f(r)}+r^2 d\Omega^2_2
\end{equation}
where $f(r)$ is given as
\begin{equation}
f(r)=1-\frac{2 G M}{r}+\frac{G Q^2 e^{-\eta }}{r^2}+\frac{r^2}{l^2}
\end{equation}
The parameter $\eta$, associated with ModMax electrodynamics, is a dimensionless quantity characterizing the theory’s nonlinearity. 
$M$ denotes the ADM mass, 
$Q$ represents the electric charge, $r$ is the horizon radius, $L$ corresponds to the AdS curvature length scale, and 
$G$ stands for Newton’s gravitational constant. By applying the condition $f(r)=0$, one can express the mass $M$ as a function of the horizon radius
\begin{equation}
M=\frac{e^{-\eta } \left(G l^2 Q^2+e^{\eta } l^2 r^2+e^{\eta } r^4\right)}{2 G l^2 r}
\end{equation}
We directly translate the mass of the ModMax-AdS black hole into the CFT energy by employing \eqref{eq:ccharge} and \eqref{eq:volume}, in conjunction with the holographic dictionary outlined in Eq. \eqref{eq:holo}
\begin{equation}
E=\frac{e^{-\eta }  \left(4 \pi  c e^{\eta } S+\pi ^2 \mathcal{Q}^2+e^{\eta } S^2\right)}{2 \pi  \sqrt{\mathcal{S}\mathcal{V}C} }
\end{equation}
The temperature on the CFT side is determined by the following expression
\begin{equation}
\mathcal{T}=\frac{e^{-\eta } \left(4 \pi  C e^{\eta } S-\pi ^2 \mathcal{Q}^2+3 e^{\eta } S^2\right)}{4 \pi  S^{3/2} \sqrt{C\mathcal{V}}}
\label{eq:temp1}
\end{equation}
The heat capacity at fixed charge is defined through the following relation
\begin{equation}
\mathcal{C}_\mathcal{Q}=\frac{2 S \left(4 \pi  C e^{\eta } S-\pi ^2 \mathcal{Q}^2+3 e^{\eta } S^2\right)}{-4 \pi  c e^{\eta } S+3 \pi ^2 \mathcal{Q}^2+3 e^{\eta } S^2}
\end{equation}
By examining the structure of the denominator, the positions of the Davies points can be identified as follows
\begin{equation}
\mathcal{S}=\frac{1}{3} \sqrt{e^{-\eta } \left(4 \pi ^2 C^2 e^{\eta }-9 \pi ^2 \mathcal{Q}^2\right)}\pm\frac{2 \pi  C}{3}
\end{equation}
When 
$\mathcal{Q}<\sqrt{\frac{4C^2}{9}e^{\eta }}$, two distinct roots appear, corresponding to the first type of Davies point. In contrast, when 
$\mathcal{Q}=\sqrt{\frac{4C^2}{9}e^{\eta }}$, these roots coincide, indicating the critical point, which is categorized as the second type of Davies point.\\
As shown in Figure \ref{fig:Fig6}, in the purple plot \ref{fig:Fig6a}, the heat capacity diverges at both types of Davies points and in the blue plot \ref{fig:Fig6b} it converges to the critical point. While the conventional Ehrenfest classification identifies these as first and second-order transitions, we later distinguish the phase structures using a generalized Ehrenfest framework.
\subsection{PHASE TRANSITION ASSOCIATED WITH THE FIRST TYPE OF DAVIES POINT}\label{sec:MODF1}
\begin{figure}[htpb]
  \centering
  \subfloat[$D^\alpha F-t$ plots]{%
    \includegraphics[width=0.3\textwidth]{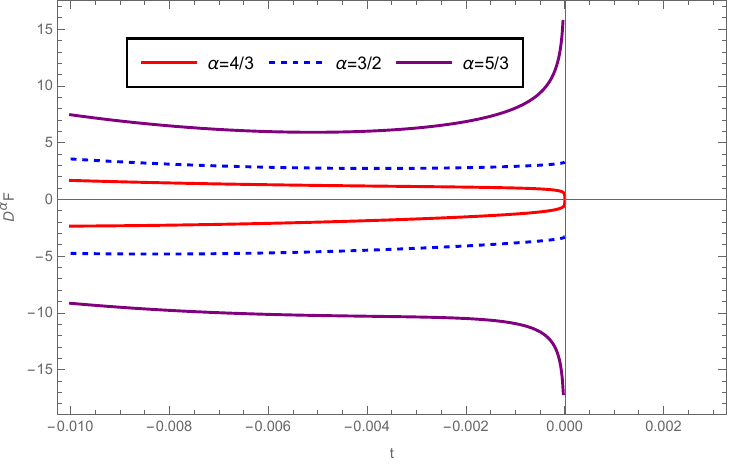}%
    \label{fig:fig1}
  }%
  \hfill
  \subfloat[$D^\alpha F-\rho$ plots]{%
    \includegraphics[width=0.3\textwidth]{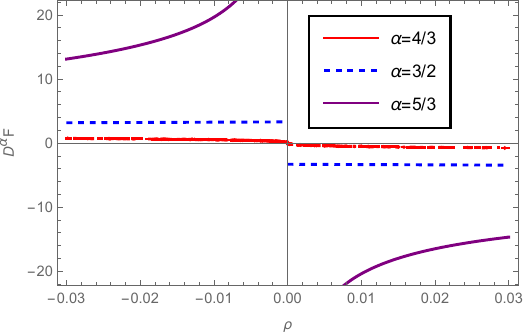}%
    \label{fig:fig2}
  }
  \caption{The behavior of 
$D^\alpha_t F(t)$ in the vicinity of the right Davies point for the CFT of ModMax AdS black hole is illustrated. The top panel shows the corresponding curves, The right bottom presents the plots of $D^\alpha_tF-\rho$ curves.}
  \label{fig:F7}
\end{figure}

\begin{figure}[htpb]
  \centering
  \subfloat[$D^\alpha F-t$ plots]{%
    \includegraphics[width=0.3\textwidth]{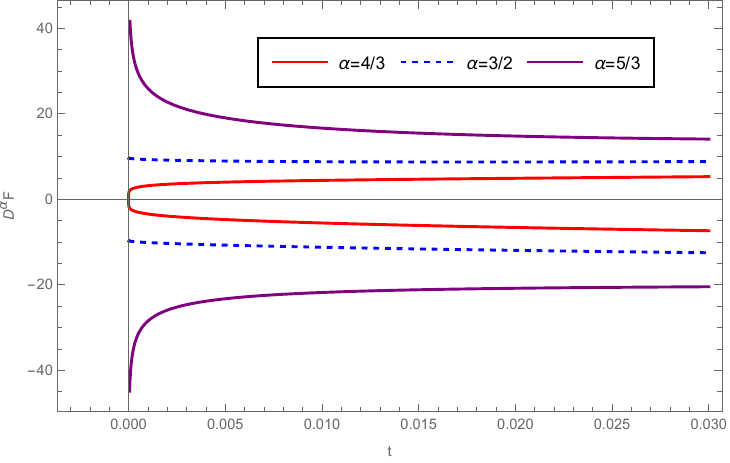}%
    \label{fig:fig1}
  }%
  \hfill
  \subfloat[$D^\alpha F-\rho$ plots]{%
    \includegraphics[width=0.3\textwidth]{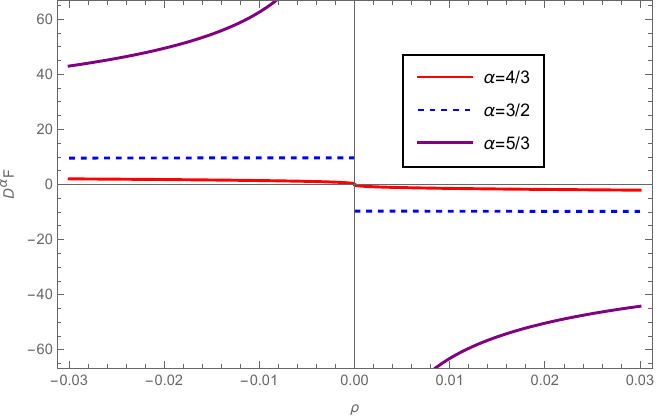}%
    \label{fig:fig2}
  }
  \caption{The behavior of 
$D^\alpha_t F(t)$ in the vicinity of the right Davies point for the CFT of Mod Max AdS black hole is illustrated. The top panel shows the corresponding curves, The right bottom presents the plots of $D^\alpha_tF-\rho$ curves.}
  \label{fig:F8}
\end{figure}

As shown in \ref{eq:temp1}, for various values of $\mathcal{S}$, the CFT specific heat of the ModMax-AdS black hole is mainly influenced by $\mathcal{Q}$, the central charge $C$, and the ModMax parameter $\gamma$. Thus, the thermal behavior near the left and right Davies points differs, requiring separate analysis of their phase structures.\\
For simplicity, we set the parameter values $\mathcal{V}=C=1$ in this section. Under this choice, the left Davies point is situated at
\begin{equation}
\mathcal{T}_c=\frac{\sqrt{\frac{3}{\pi }} \left(4 e^{\eta }-2 \sqrt{e^{\eta } \left(4 e^{\eta }-9 \mathcal{Q}^2\right)}-3 \mathcal{Q}^2\right)}{\left(2 e^{\eta }-\sqrt{e^{\eta } \left(4 e^{\eta }-9 \mathcal{Q}^2\right)}\right) \sqrt{2-e^{-\eta } \sqrt{e^{\eta } \left(4 e^{\eta }-9 \mathcal{Q}^2\right)}}}
\end{equation}
\begin{equation}
\mathcal{S}_c=\frac{1}{3} \pi  \left(2-e^{-\eta } \sqrt{e^{\eta } \left(4 e^{\eta }-9 \mathcal{Q}^2\right)}\right)
\end{equation}
We adopt the dimensionless variables defined in \eqref{eq:dim}. In this framework, the Davies point is located at 
$t=\rho=0$. This facilitates a series expansion in the neighbourhood of the Davies point. As the left Davies point corresponds to the maximum of the temperature profile, the condition 
$t\leq 0$ holds, consistently representing the highest attainable temperature.\\
By choosing 
$\mathcal{Q}=1$ and $\gamma=1$, a dimensionless version of the equation of state can be obtained from the original expression in Eq. \eqref{eq:temp1}
\begin{equation}
-\frac{1.081 \rho ^2}{(\rho +1)^{3/2}}-\frac{5.861 \rho }{(\rho +1)^{3/2}}-\frac{3.907}{(\rho +1)^{3/2}}+3.907 t+3.907=0
\label{eq:eos2}
\end{equation}
In the neighbourhood of the Davies point, we solve \eqref{eq:eos2} to obtain a series expansion for $\rho$ expressed as a function of the variable $t$
\begin{equation}
\begin{split}
&\rho = -4924.851 t^3+191.589 t^2 -42.847 (- t)^{3/2}\\
& -940.161 (- t)^{5/2} -10.878 t  -3.191 \sqrt{- t} \\
&\rho = -4924.851 t^3 +191.589 t^2 +42.847 (- t)^{3/2}\\
& +940.161 (- t)^{5/2} -10.878 t  +3.191 \sqrt{- t} 
\end{split}
\label{eq:rho2}
\end{equation}
The two series solutions correspond to separate branches around the Davies point, notably featuring terms with fractional exponents. The free energy is derived using the dimensionless quantities and is expressed as follows
\begin{equation}
\label{eq:free2}
F=\frac{A_1}{12 \sqrt{-e^{-\eta } (\rho +1) \left(\sqrt{e^{\eta } \left(4 e^{\eta }-9 \mathcal{Q}^2\right)}-2 e^{\eta }\right)}}
\end{equation}
\begin{equation*}
\begin{split}
&A_1=\sqrt{\frac{\pi }{3}} e^{-\eta } \left(-8 e^{\eta } \left(\rho ^2-\rho -2\right)\right.\\&\left.+4 \left(\rho ^2-\rho -2\right) \sqrt{e^{\eta } \left(4 e^{\eta }-9 \mathcal{Q}^2\right)}+9 \left(\rho ^2+2 \rho +4\right) \mathcal{Q}^2\right)
\end{split}
\end{equation*}
By inserting the solutions obtained from \eqref{eq:rho2} into the free energy expression \eqref{eq:free2}, we arrive at the following result
\begin{equation}
\begin{split}
&F=-16293.7 t^4 -74.689 t^3 +6.360 t^2 -2.488 (- t)^{3/2}\\
&-20.0441 (- t)^{5/2} -322.345 (- t)^{7/2} -1.169 t  +1.782\\
&F=-16293.7 t^4 -74.689 t^3 +6.360 t^2 +2.488 (-t)^{3/2}\\
&+20.0441 (-t)^{5/2} +322.345 (-t)^{7/2} -1.169t +1.782
\end{split}
\end{equation}
It is noteworthy that the expansion contains terms with fractional exponents. In particular, the coefficients preceding the $(-t)^{3/2}$ term have opposite signs.\\
We now proceed to compute the fractional derivatives of the free energy using Caputo’s definition. Owing to the condition 
$t \leq 0$, the 
$D^\alpha_t F-t$ curves appear on the left side of the 
$t=0$ axis. Alternatively, expressing 
$D^\alpha_t F-t$ as a function of $\rho$ allows its behavior to manifest on both sides of the 
$\rho=0$ axis. By substituting specific values of $\alpha$ into \eqref{eq:series} and analyzing the limiting behavior as $t\rightarrow 0$, we obtain the following result
\[
\lim_{t \to 0^-} D_t^\alpha F(t) =
\begin{cases}
0, & \alpha < 3/2, \\
 \pm 3.307  , & \alpha =3/2, \\
\pm \infty, & \alpha > 3/2.
\end{cases}
\]
As shown in Figure \ref{fig:F7}, the behavior of 
$D^\alpha_t F(t)$ reveals continuity at the Davies point for $\alpha<3/2$, a finite jump at 
$\alpha =3/2$, and divergence for 
$\alpha>3/2$. According to the generalized classification, this signifies a phase transition of order 
$\alpha =3/2$ at the Davies point.\\
We now turn to the analysis of the phase structure of the CFT dual to the ModMax-AdS black hole in the vicinity of the right Davies point. The approach adopted here closely mirrors that of the preceding investigation. In this context, the Davies point is positioned at
\begin{equation}
\begin{split}
&\mathcal{T}_c=\frac{\sqrt{\frac{3}{\pi }} \left(4 e^{\eta }+2 \sqrt{e^{\eta } \left(4 e^{\eta }-9 \mathcal{Q}^2\right)}-3 \mathcal{Q}^2\right)}{\left(2 e^{\eta }+\sqrt{e^{\eta } \left(4 e^{\eta }-9 \mathcal{Q}^2\right)}\right) \sqrt{e^{-\eta } \sqrt{e^{\eta } \left(4 e^{\eta }-9 \mathcal{Q}^2\right)}+2}}\\
&\mathcal{S}_c=\frac{1}{3} \pi  \left(e^{-\eta } \sqrt{e^{\eta } \left(4  e^{\eta }-9 \mathcal{Q}^2\right)}+2\right)
\end{split}
\end{equation}
By employing the dimensionless variables defined in \eqref{eq:dim}, the equation of state reduces to the following simplified form
\begin{equation}
\begin{split}
&\frac{0.077 \left(3 \left(3 \rho ^2+6 \rho +4\right)-30.772 \left(\rho ^2+3 \rho +2\right)\right)}{(\rho +1)^{3/2}}\\
&+3.828 (t +1)=0
\end{split}
\label{eq:eos3}
\end{equation}
We solve  \eqref{eq:eos3} to obtain 
$\rho(t)$ as a series expansion, and then substitute the resulting expression into the free energy formula \eqref{eq:free2}. This leads to the following form of the free energy
\begin{equation}
\begin{split}
&F=7.278 t^{3/2} +9.986 t^{5/2} +16.701 t^3 -5.700 t^2\\
& -2.771 t  +1.819\\
&F=-7.278 t^{3/2} -9.986 t^{5/2} +16.701 t^3 -5.700 t^2\\
& -2.771 t +1.819
\end{split}
\end{equation}
Apart from similarities in coefficient values and the variable $t$, the functional forms of these free energy expressions closely resemble those near the left Davies point, resulting in similar behavior for their fractional derivatives. Figure \ref{fig:F8} illustrates the behavior of these derivatives as functions of $t$ and $\rho$, with their limiting values provided below
\[
\lim_{t \to 0^+} D_t^\alpha F(t) =
\begin{cases}
0, & \alpha < 3/2, \\
 \pm 9.675  , & \alpha =3/2, \\
\pm \infty, & \alpha > 3/2.
\end{cases}
\]
In summary, although the temperature profiles differ between the left and right Davies points, the underlying phase structures at both locations exhibit similar characteristics. Remarkably, the phase transitions at these points are of fractional order, specifically $\alpha =3/2$ rather than the conventionally expected first order. To the best of our knowledge, such non-monotonic thermal behaviour particularly that observed near the first type of Davies point is uncommon in standard thermodynamic systems when analysed within the CFT thermodynamic framework. While similar features have been explored on the gravitational side for various black hole solutions, their manifestation in the dual CFT remains relatively less studied and warrants further investigation.

\subsection{PHASE TRANSITION ASSOCIATED WITH THE SECOND TYPE OF DAVIES POINT}\label{sec:MODF2}
We now focus on the second class of Davies points, namely the critical points, which are specified as follows
\begin{equation}
\mathcal{Q}_c= \frac{2}{3} C e^{\eta /2},\quad \mathcal{S}_c= \frac{2 \pi  C}{3},\quad \mathcal{T}_c=\frac{2 \sqrt{\frac{2}{3 \pi }}}{\sqrt{\mathcal{V}}}
\end{equation}
Utilizing the definition presented in \eqref{eq:dim}, we likewise define 
$\mathcal{Q}=\mathcal{Q}_c(1+q)$. Under this substitution, the equation of state simplifies to the following expression
\begin{equation}
\begin{split}
&0.115 \left(q^2+2q+8 t\right)-0.172 \left(q^2+2 q\right) \rho \\
&+0.215 \left(q^2+2 q\right) \rho ^2-0.251 \left(q^2+2 q+0.228\right) \rho ^3\\
&+0.283 \left(q^2+2 q+0.380\right) \rho ^4=0
\end{split}
\end{equation}
This represents a quartic equation in the variable 
$\rho$, whose real roots are provided as follows
\begin{equation}
\begin{split}
&\rho (t)=3.968 t^{2/3} +4.593 t^{4/3}+3.654 t^{5/3}+1.186 t^{7/3}\\
& +0.385t^{8/3} +0.001 t^3 +2.367 t^2+2.519 t^{1/3} +4.75 t 
\end{split}
\label{eq:rho3}
\end{equation}
It is essential to highlight that this series expansion holds true exclusively within the domain defined by the condition $2q+8t\neq 0$\\
In the neighbourhood of the critical point, the normalized free energy assumes the following expression.
\begin{equation}
\tilde{F}=F/F_c=\frac{ \left(3 q^2-\rho ^2+6 q+4 \rho +8\right)}{8 \sqrt{ (\rho +1)}}
\end{equation}
where $F_c=\frac{4 \sqrt{\frac{2 \pi }{3}} C}{3 \sqrt{\mathcal{V}}}$. By substituting \eqref{eq:rho3} into the
  expression for the reduced free energy, we obtain the following result
  \begin{equation}
  \begin{split}
& \tilde{F}(q,t)= 0.297 q t^{2/3}+(-0.009 q-1.889) t^{4/3}\\
 &+(-0.031 q-2.381) t^{5/3}+(-0.014 q-2.375) t^2\\
 &-0.944 q t^{1/3}+(0.093q-1) t+\frac{3 q}{4}+1
 \end{split}
 \label{eq:free1}
  \end{equation}
  Next, we compute the fractional derivatives of the free energy and investigate their behavior in the limiting case as 
$t\rightarrow 0^+$. The corresponding expressions are provided below.
\[
\lim_{t \to 0^+} D_t^\alpha \tilde{F}(t) =
\begin{cases}
0, & \alpha < 4/3, \\
  9.00067  , & \alpha =4/3, \\
\pm \infty, & \alpha > 4/3.
\end{cases}
\]
It is noted that the fractional derivative of the free energy of order $4/3$ displays a discontinuity, indicating a hallmark signature of a phase transition of the same order. By considering the limit 
$q\rightarrow 0$, which corresponds to the critical point in \eqref{eq:free1}, we obtain the following
\[
F(t) =
\begin{cases}
7.55953 t^{4/3}+26.7874 t^{5/3}+86.125 t^2\\+ t+1, & t>0,  \\
-1.88988 t^{4/3}-2.3811 t^{5/3}-2.375 t^2\\-t+1, & t<0.
\end{cases}
\]
Owing to the presence of the 
$t^{4/3}$ terms, a finite discontinuity 
arises at $\alpha=4/3$
\begin{equation}
\lim_{t\rightarrow 0^-}\frac{d^\alpha \tilde{F}}{dt^\alpha}=-\frac{3.0469}{\Gamma \left(\frac{2}{3}\right)}\neq \lim_{t\rightarrow0^+}\frac{d^\alpha \tilde{F}}{dt^\alpha}=\frac{12.1888}{\Gamma \left(\frac{2}{3}\right)}
\end{equation}
According to the generalized classification framework, the phase transition at the critical point is characterized as being of fractional order 
$\alpha=4/3$. The associated behaviour is depicted in Figure \ref{fig:F9}. 
\begin{figure}
\includegraphics[scale=0.5]{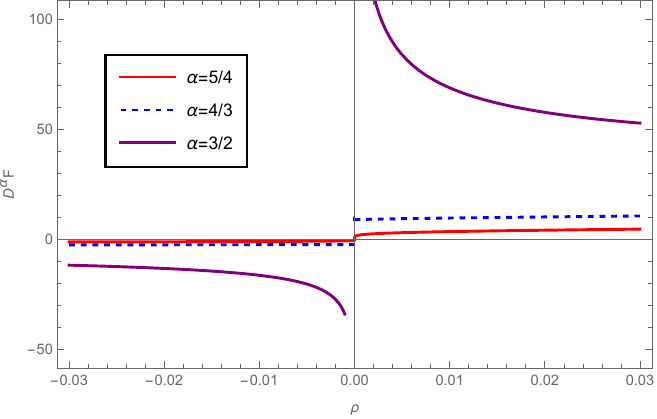}
\caption{The behavior of \( D^\alpha_t \tilde{F}(\rho) \) in the vicinity of the critical point is examined within the CFT thermodynamic framework corresponding to the ModMax-AdS black hole.
}
\label{fig:F9}
\end{figure}
Furthermore, it is essential to consider the exceptional scenario defined by the constraint 
$2t+8q=0$, which delineates a distinct trajectory within the 
$(t,q)$ parameter space. In this case, the critical point must be approached precisely along this specific path from both directions. Under this imposed condition, the equation of state simplifies to the following reduced form
\begin{equation}
\begin{split}
&0.283 \rho ^4 \left(16 t^2-8t+0.380\right)-0.251 \rho ^3 \left(16 t^2-8 t+0.228\right)\\
&+0.215 \rho ^2 \left(16t^2-8 t\right)-0.172 \rho  \left(16. t^2-8t\right)+1.842 t^2=0
\end{split}
\end{equation}
The associated series expanded solutions are presented below in the following expressions
\begin{equation}
\begin{split}
&\rho=-160.433 t^{3/2}-21324.1 t^{5/2}+300652. t^3+1462.77 t^2\\
&+8.166 t-4.898 \sqrt{t}\\
&\rho=160.433 t^{3/2}+21324.1 t^{5/2}+300652. t^3+1462.77 t^2\\
&+8.166 t+4.898 \sqrt{t}
\end{split}
\end{equation}
In this particular scenario, the appearance of terms with fractional powers leads to a reduced free energy expression that also incorporates fractional exponents. Notably, such behavior was absent in the RN-AdS black hole case. The dimensionless free energy corresponding to the condition 
$2t+8q=0$ is given by
\[
F(t) =
\begin{cases}
7.34847 t^{3/2}+711.564 t^{5/2}-3751.76 t^3\\+3.75 t^2+1, & t>0,  \\
-7.34847 t^{3/2}-711.564 t^{5/2}-3751.76 t^3\\+3.75 t^2+1, & t<0.
\end{cases}
\]
We evaluate the fractional derivatives of the free energy and investigate their behavior in the limiting cases as 
$t\rightarrow 0^+$ and 
$t\rightarrow 0^-$ . The corresponding analytical expressions are presented below
\[
\lim_{t \to 0^-} D_t^\alpha F(t) =
\begin{cases}
0, & \alpha < 3/2, \\
 \pm 9.76862  , & \alpha =3/2, \\
\pm \infty, & \alpha > 3/2.
\end{cases}
\]
As shown above and illustrated in Figure 10, the fractional derivative of the free energy of order $\alpha =3/2$ exhibits a finite discontinuity, signalling the presence of a phase transition of order $3/2$ under the special condition 
$2t+8q=0$. This characteristic behavior is notably absent in the analogous RN-AdS black hole case.\\
\begin{figure}
\includegraphics[scale=0.5]{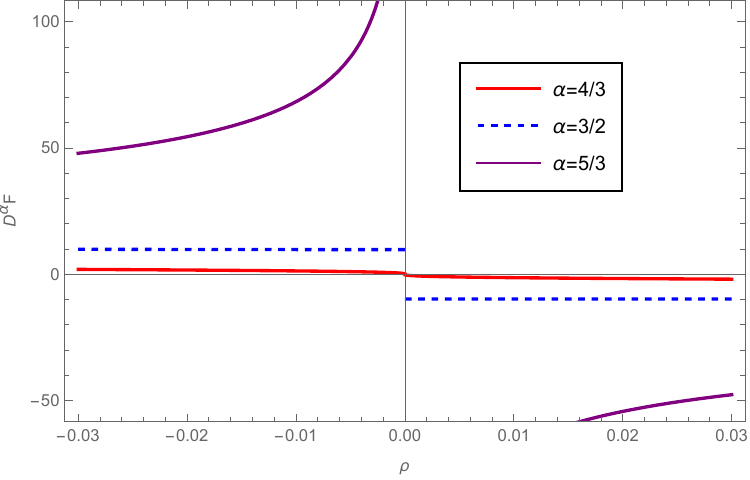}
\caption{The behavior of \( D^\alpha_t \tilde{F}(\rho) \) in the vicinity of the critical point is examined within the CFT thermodynamic framework corresponding to the ModMax-AdS black hole, particularly along a special direction in the parameter space.}
\label{fig:F9}
\end{figure}
The appearance of different fractional orders $4/3$ in the general case and 
$3/2$ along the fine-tuned direction 
$2q+8t=0$ reflects distinct pathways to the critical point in thermodynamic parameter space. The 
$4/3$ order transition arises generically and captures the dominant critical behavior, thereby representing the physically significant classification. In contrast, the 
$3/2$ order transition emerges only under specific conditions and should be interpreted as a special-case phenomenon. Hence, the 
$4/3$ order transition is identified as the principal result.
\section{RN-ADS BLACK HOLE WITH KANIADAKIS STATISTICS}\label{sec:KAN}
The metric of RN-ADS is derived in \ref{sec:RNADS} and expressed in the following form
\begin{equation}
ds^2=-f(r)dt^2+\frac{1}{f(r)}dr^2+r^2 d\theta^2+r^2 \sin^2\theta d\phi^2
\end{equation}
where $f(r)=1+\frac{r^2}{l^2}-\frac{2G M}{r}+\frac{G Q^2}{r^2}$. By enforcing the condition 
$f(r)=0$, the mass $M$ can be determined as a function of the horizon radius as
\begin{equation}
M=\frac{G l^2 Q^2+l^2 r^2+r^4}{2 G l^2 r}
\end{equation}
In the framework of Kaniadakis statistics, the entropy is expressed as follows
\begin{equation}
S=\frac{1}{\kappa}\sinh \left(\frac{\pi \kappa r^2}{G}\right)
\end{equation}
Here, $\kappa$ denotes the deformation parameter, which characterizes the deviation from the conventional Boltzmann–Gibbs statistical formalism. We translate the mass of the ModMax-AdS black hole into the corresponding CFT energy by applying \eqref{eq:ccharge} and \eqref{eq:volume}, along with the holographic dictionary outlined in \eqref{eq:holo}, and by solving the Kaniadakis entropy expression above to determine the value of $r$ in terms of $\mathcal{S}$ and putting it back in the mass equation. We series expand the CFT energy for small values of $\kappa$ hence we get
\begin{equation}
E=\frac{A}{24 \pi  C \sqrt{\frac{\mathcal{S} \mathcal{V}}{C}}}
\end{equation}
where $A=-4 \pi  C \mathcal{S} \left(\kappa ^2 \mathcal{S}^2-12\right)+12 \pi ^2 \mathcal{Q}^2-3 \kappa ^2 \mathcal{S}^4+\mathcal{S}^2 \left(\pi ^2 \kappa ^2 \mathcal{Q}^2+12\right)$.
The temperature on the CFT side is computed using the following expression
\begin{equation}
\mathcal{T}=\frac{B}{48 \pi  C^2 \left(\frac{\mathcal{S} \mathcal{V}}{C}\right)^{3/2}}
\label{eq:temp2}
\end{equation}
\begin{equation}
\begin{split}&B=\mathcal{V} \left(4 \pi  C \mathcal{S} \left(12-5 \kappa ^2 \mathcal{S}^2\right)\right.\\&\left.+3 \left(-4 \pi ^2 \mathcal{Q}^2-7 \kappa ^2 \mathcal{S}^4+\mathcal{S}^2 \left(\pi ^2 \kappa ^2 \mathcal{Q}^2+12\right)\right)\right)\end{split}
\end{equation}
The heat capacity at fixed charge is characterized by the following expression
\begin{equation}
\mathcal{C}_\mathcal{Q}=\frac{D}{E}
\end{equation}where
\begin{equation}
\begin{split}
&D=2 S \left(4 \pi  C \mathcal{S} \left(5 \kappa ^2 \mathcal{S}^2-12\right)+12 \pi ^2 \mathcal{Q}^2+21 \kappa ^2 \mathcal{S}^4\right.\\&\left.-3 \mathcal{S}^2 \left(\pi ^2 \kappa ^2 \mathcal{Q}^2+12\right)\right)\\
& E=3\left(4 \pi  C \mathcal{S} \left(5 \kappa ^2 \mathcal{S}^2+4\right)-12 \pi ^2 \mathcal{Q}^2+35 \kappa ^2 \mathcal{S}^4\right.\\&\left.-\mathcal{S}^2 \left(\pi ^2 \kappa ^2 \mathcal{Q}^2+12\right)\right)
\end{split}
\end{equation}
By analyzing the form of the denominator $E$, the locations of the Davies points can be determined as follows
\begin{equation}
\begin{split}
&\mathcal{S}=\text{Root}\left[35 \text{$\#$1}^4 \kappa ^2+20 \pi  \text{$\#$1}^3 C \kappa ^2+\text{$\#$1}^2 \left(-\pi ^2 \kappa ^2 \mathcal{Q}^2-12\right)\right.\\&\left.+16 \pi  \text{$\#$1} C-12 \pi ^2 \mathcal{Q}^2\&,2\right]\\
&\mathcal{S}=\text{Root}\left[35 \text{$\#$1}^4 \kappa ^2+20 \pi  \text{$\#$1}^3 C \kappa ^2+\text{$\#$1}^2 \left(-\pi ^2 \kappa ^2 \mathcal{Q}^2-12\right)\right.\\&\left.+16 \pi  \text{$\#$1} C-12 \pi ^2 \mathcal{Q}^2\&,3\right]\\
& \mathcal{S}=\text{Root}\left[35 \text{$\#$1}^4 \kappa ^2+20 \pi  \text{$\#$1}^3 C \kappa ^2+\text{$\#$1}^2 \left(-\pi ^2 \kappa ^2 \mathcal{Q}^2-12\right)\right.\\&\left.+16 \pi  \text{$\#$1} C-12 \pi ^2 \mathcal{Q}^2\&,4\right]
\end{split}
\end{equation}
It is observed that the use of Kaniadakis entropy results in the appearance of three distinct Davies points. This represents a unique characteristic of adopting the Kaniadakis statistical framework
\begin{figure}
\includegraphics[scale=0.7]{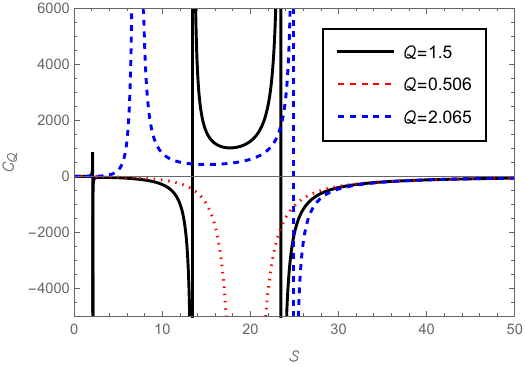}
\caption{The heat capacity for the CFT dual of the RN-AdS black hole, incorporating Kaniadakis entropy, is depicted as a function of the horizon radius. The parameters are suitably set to $C=3$, $\kappa=0.016$. The black, red-dotted and blue dashed curves correspond to the cases with 
$\mathcal{Q}=1.5$, $ \mathcal{Q}=0.506$, $\mathcal{Q}=2.065$ respectively.}
\label{fig:F11}
\end{figure}
As illustrated in Figure \ref{fig:F11}, the heat capacity in $\mathcal{\
Q}=1.5$ (black curve) exhibits divergences at three types of Davies points, whereas in $\mathcal{Q}=0.506$ (red-dotted curve), it converges towards the first critical point and at $\mathcal{Q}=2.065$ (blue-dashed curve), it converges to the second critical point. Although the traditional Ehrenfest classification designates these as second-order phase transitions, we subsequently differentiate the phase structures using a generalized Ehrenfest framework.
\subsection{PHASE TRANSITION ASSOCIATED WITH THE FIRST TYPE OF DAVIES POINT}\label{sec:KANF1}
\begin{figure}[htpb]
  \centering
  \subfloat[$D^\alpha F-t$ plots]{%
    \includegraphics[width=0.3\textwidth]{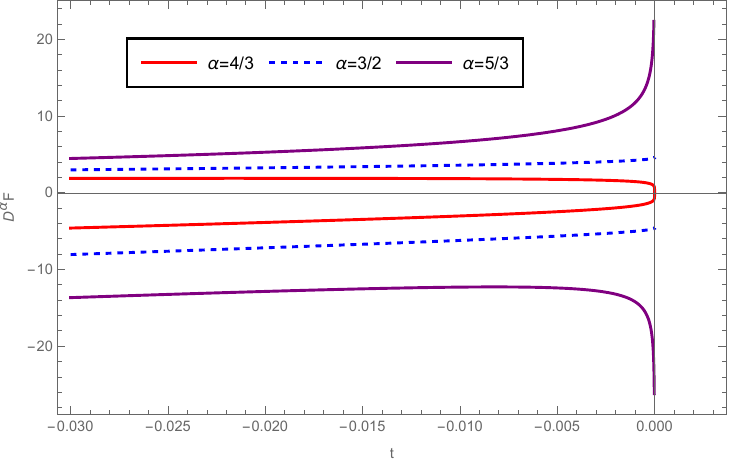}%
    \label{fig:fig1}
  }%
  \hfill
  \subfloat[$D^\alpha F-\rho$ plots]{%
    \includegraphics[width=0.3\textwidth]{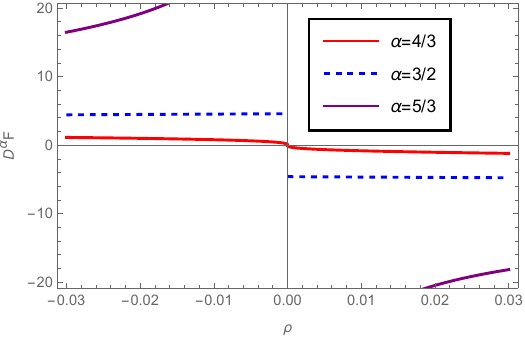}%
    \label{fig:fig2}
  }
  \caption{The behaviour of \( D^\alpha_t F(t) \) in the vicinity of the first Davies point is illustrated within the CFT thermodynamic framework corresponding to the RN-AdS black hole formulated using Kaniadakis statistics. The top panel shows the corresponding \( D^\alpha_t F(t) \) vs. \( t \) curves, while the bottom right panel presents the plots of \( D^\alpha_t F \) vs. \( \rho \).
}
  \label{fig:F12}
\end{figure}

\begin{figure}[htpb]
  \centering
  \subfloat[$D^\alpha F-t$ plots]{%
    \includegraphics[width=0.3\textwidth]{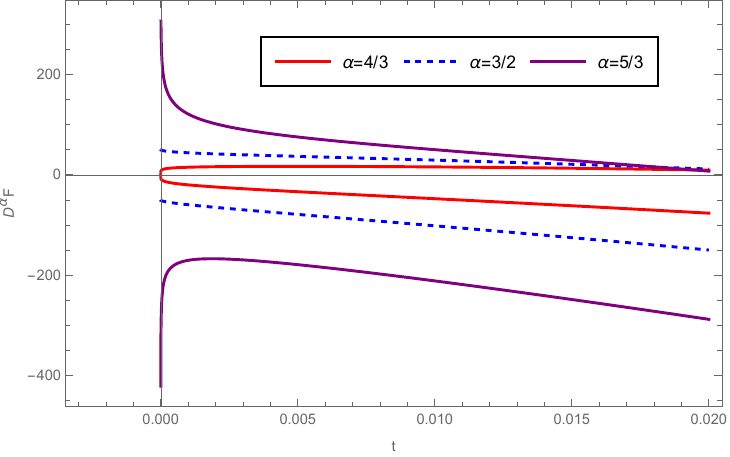}%
    \label{fig:fig1}
  }%
  \hfill
  \subfloat[$D^\alpha F-\rho$ plots]{%
    \includegraphics[width=0.3\textwidth]{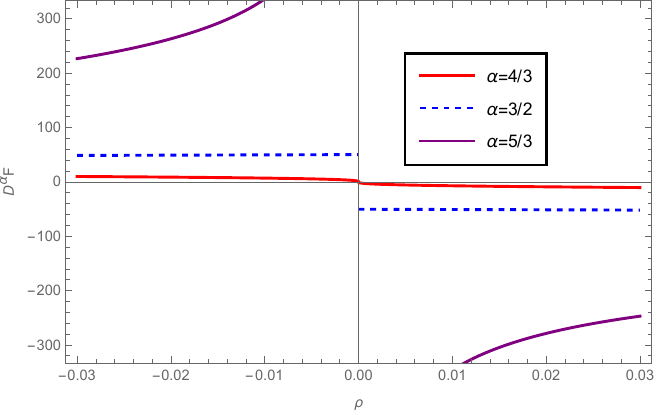}%
    \label{fig:fig2}
  }
  \caption{The behavior of 
$D^\alpha_t F(t)$ in the vicinity of the right Davies point for the CFT of Mod Max AdS black hole is illustrated. The top panel shows the corresponding curves, The right bottom presents the plots of $D^\alpha_tF-\rho$ curves.}
  \label{fig:F13}
\end{figure}

As indicated in \eqref{eq:temp2}, for different values of 
$\mathcal{S}$, the specific heat on the CFT side of the RN AdS black hole incorporated with the Kaniadakis entropy is predominantly influenced by the charge $\mathcal{Q}$, the central charge $C$, and the kaniadakis parameter $\kappa$. Consequently, the thermal behavior near the first, second and third Davies points varies, necessitating an independent analysis of their respective phase structures.\\
For simplicity, the parameter values are fixed as 
$\mathcal{V}=1$, $C=3$, $\mathcal{Q}=1.5$ and $\kappa=0.016$ in this section. With this choice, the first Davies point is located at the following position
\begin{equation}
\mathcal{T}_c=1.05913,\quad \mathcal{S}_c=2.12107
\end{equation}
We employ the dimensionless variables introduced in \eqref{eq:dim}. Within this framework, the Davies point is positioned at 
$t=\rho=0$, enabling a series expansion to be carried out in the vicinity of the Davies point. By setting 
$\mathcal{Q}=1.5$, $C=3$, $\mathcal{V}=1$ and 
$\kappa=0.016$, a dimensionless form of the equation of state can be derived from the original relation presented in \eqref{eq:temp2} using \eqref{eq:dim}
\begin{equation}\begin{split}
&1.059 (t  +1)\\&-\frac{-0.0001 \rho ^4-0.0011 \rho ^3+0.198 \rho ^2+1.588 \rho +1.05}{\sqrt{\rho +1} (\rho +1)}=0\end{split}
\label{eq:eos4}
\end{equation}
In the vicinity of the Davies point, \eqref{eq:eos4} is solved to obtain a series expansion of $\rho$ expressed as a function of the variable $t$.
\begin{equation}
\begin{split}
&\rho=-119.923 t^3 +22.646 t^2 -10.339 (- t)^{3/2}-51.313 (-t)^{5/2}\\
& -4.866 t  -2.308 \sqrt{-t}\\
&\rho= -119.923 t^3 +22.646 t^2 +10.339 (-1t)^{3/2} +51.313 (- t)^{5/2}\\
& -4.866 t  +2.308 \sqrt{- t}  
\end{split}
\label{eq:rho}
\end{equation}
The two series solutions represent distinct branches near the Davies point and are characterized by the presence of terms with fractional powers. The free energy is formulated using the dimensionless variables and is expressed as follows.
\begin{equation}
\begin{split}
&F=\frac{1}{\sqrt{\rho +1}}\left(
\begin{aligned}
&0002 \rho ^4+0.0015 \rho ^3-0.138 \rho ^2+2.241 \rho\\&+\left(-0.00002 \rho ^2-0.00005 \rho +0.934\right) \mathcal{Q}^2\\&+2.381
\end{aligned}\right)
\end{split}
\end{equation}
By substituting the solutions derived from \eqref{eq:eos4} into the free energy expression, we obtain the following result
\begin{equation}
\begin{split}
&F=-16.957 t^3 +5.465 t^2 -3.456 (-t)^{3/2} -9.290 (- t)^{5/2}\\
&-2.246 t  +4.483\\
&F=-16.957 t^3 +5.465 t^2 +3.456 (- t)^{3/2} +9.290 (-t)^{5/2}\\
& -2.246 t  +4.483
\end{split}
\end{equation}
It is noteworthy that the series expansion includes terms with fractional exponents. Specifically, the coefficients preceding the 
$(-t)^{3/2}$
  term exhibit opposite signs.\\
  We now proceed to evaluate the fractional derivatives of the free energy employing Caputo’s formalism. Due to the condition 
$t\leq 0$, the curves corresponding to 
$D^\alpha_tF-t$  emerge on the left side of the 
$t=0$ axis. Alternatively, expressing 
$D^\alpha_tF-t$ as a function of $\rho$ enables its behavior to be analyzed on both sides of the 
$\rho=0$ axis. By substituting specific values of $\alpha$ into \eqref{eq:series} and examining the limiting behavior at 
$t$, we arrive at the following result.
\[
\lim_{t \to 0^-} D_t^\alpha F(t) =
\begin{cases}
0, & \alpha < 3/2, \\
 \pm 4.5946   , & \alpha =3/2, \\
\pm \infty, & \alpha > 3/2.
\end{cases}
\]
As illustrated in Figure \ref{fig:F12}, 
the behaviour of $D^\alpha_tF(t)
$ exhibits continuity at the 
Davies point for $\alpha <3/2$, 
a finite discontinuity at $\alpha=3/2$
 , and divergence for 
$\alpha >3/2$ . In accordance with the generalized classification scheme, this indicates a phase transition of order 
$\alpha=3/2$
  occurring at the first Davies point.\\
  We now shift our focus to analyzing the phase structure of the CFT dual to the RN-AdS black hole within the framework of Kaniadakis entropy, specifically in the vicinity of the second Davies point. The methodology employed here closely parallels that of the previous analysis. In this setting, the Davies point is located at the following position
  \begin{equation}
  \mathcal{T}_c=0.934556, \quad \mathcal{S}_c=13.4586
  \end{equation}
  By utilizing the dimensionless variables introduced in \eqref{eq:dim}, the equation of state simplifies to the following reduced form
  \begin{equation}\begin{split}
 & 0.934 (t+1)\\&-\frac{-0.013 \rho ^4-0.063 \rho ^3+0.396 \rho ^2+1.401\rho +0.934}{\sqrt{\rho +1} (\rho +1)}=0
  \end{split}
  \label{eq:eos5}
  \end{equation}
We solve \eqref{eq:eos5} to derive $\rho (t)$ as a series expansion, and subsequently substitute the resulting expression into the free energy formulation \eqref{eq:free3}. This yields the following representation of the free energy
\begin{equation}
\begin{split}
& F(t)=37.795 t^{3/2} +459.234 t^{5/2} -2833.99 t^3 -103.355 t^2\\
& -12.577 t  +5.124\\
&F(t)=-37.795 t^{3/2} -459.234 t^{5/2} -2833.99 t^3-103.355 t^2\\
& -12.577 t  +5.124
\end{split}
\label{eq:free3}
\end{equation}
Aside from similarities in the coefficient values and the variable $t$, the functional forms of these free energy expressions closely parallel those observed near the first Davies point. Consequently, their fractional derivatives exhibit analogous behaviour. Figure \ref{fig:F13} presents the behavior of these derivatives as functions of both 
$t$ and $\rho$, with the corresponding limiting values listed below
\[
\lim_{t \to 0^-} D_t^\alpha F(t) =
\begin{cases}
0, & \alpha < 3/2, \\
 \pm 50.2425   , & \alpha =3/2, \\
\pm \infty, & \alpha > 3/2.
\end{cases}
\]
In conclusion, while the temperature profiles differ between the first and second Davies points, the underlying phase structures at both locations display similar features. Notably, the associated phase transitions are of fractional order, specifically $\alpha=3/2$ , rather than the traditionally anticipated first-order transitions.\\
We now turn to the investigation of the phase structure of the CFT dual to the RN-AdS black hole, formulated within the context of Kaniadakis entropy, particularly near the third Davies point. The analytical approach adopted here closely mirrors the preceding methodology. Under this framework, the Davies point is situated at the following location.
\begin{equation}
\mathcal{T}_c=0.940672,\quad \mathcal{S}_c=23.4543
\end{equation}
By utilizing the dimensionless parameters introduced in Eq.
 \eqref{eq:dim}, the equation of state simplifies to the following reduced form
 \begin{equation}
 \begin{split}
& 0.940 (t+1)\\&-\frac{-0.054 \rho ^4-0.240 \rho ^3+0.275 \rho ^2+1.411 \rho +0.940}{(\rho +1)^{3/2}}=0
 \end{split}
 \label{eq:eos6}
 \end{equation}
 We solve \eqref{eq:eos6} to derive a series expansion for 
$\rho (t)$, and subsequently insert this expression into the free energy formulation. This procedure yields the following expression for the free energy
\begin{equation}
\begin{split}
&F=1977.9 t^3 -58.0948 t^2 -51.4457 (-t)^{3/2} -300.793 (-t)^{5/2}\\
& -22.0628 t  +5.01307\\
&F= 1977.9 t^3 -58.0948 t^2 +51.4457 (-t)^{3/2} +300.793 (-t)^{5/2}\\
& -22.0628 t  +5.01307
\end{split}
\end{equation}
Aside from similarity in the coefficient magnitudes and the parameter 
$t$, the structural form of these free energy expressions closely aligns with those found near the left Davies point. Consequently, their fractional derivatives exhibit comparable behavior. Figure 12 displays the evolution of these derivatives as functions of 
$t$ and 
$\rho$, with the associated limiting values detailed below
\[
\lim_{t \to 0^+} D_t^\alpha F(t) =
\begin{cases}
0, & \alpha < 3/2, \\
 \pm 68.3889   , & \alpha =3/2, \\
\pm \infty, & \alpha > 3/2.
\end{cases}
\]
\begin{figure}[htpb]
  \centering
  \subfloat[$D^\alpha F-t$ plots]{%
    \includegraphics[width=0.3\textwidth]{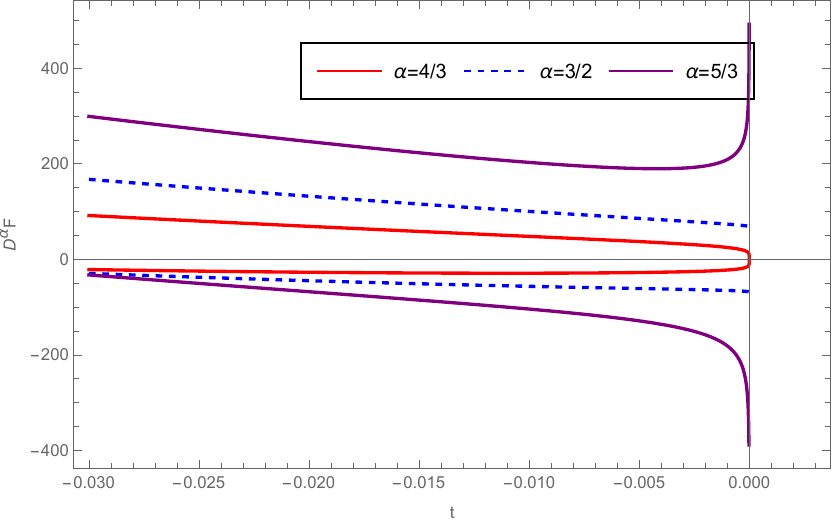}%
    \label{fig:fig1}
  }%
  \hfill
  \subfloat[$D^\alpha F-\rho$ plots]{%
    \includegraphics[width=0.3\textwidth]{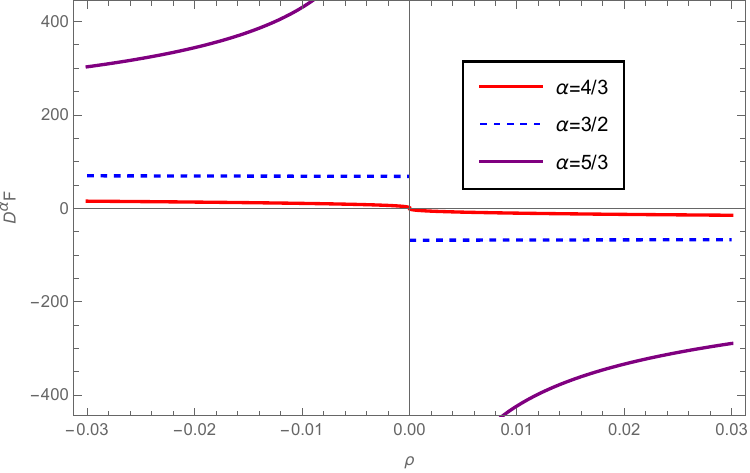}%
    \label{fig:fig2}
  }
  \caption{The behavior of 
$D^\alpha_t F(t)$ in the vicinity of the right Davies point for the CFT of Mod Max AdS black hole is illustrated. The top panel shows the corresponding curves, The right bottom presents the plots of $D^\alpha_tF-\rho$ curves.}
  \label{fig:sidebyside}
\end{figure}
In conclusion, despite the distinct temperature profiles observed at the first, second and third Davies points, the corresponding phase structures at both locations reveal analogous features. Notably, the associated phase transitions are of fractional order, characterized by 
$\alpha =3/2$, rather than being first-order as traditionally anticipated. 
\subsection{ PHASE TRANSITION ASSOCIATED WITH THE SECOND TYPE OF DAVIES POINT}\label{sec:KANF2}
 We now direct our attention to the second category of Davies points, referred to as critical points, which are identified by the following conditions
\begin{equation}
\mathcal{T}_c=0.949,\quad \mathcal{S}_c=19.339,\quad \mathcal{Q}_c=0.506
\end{equation}
Employing the definition introduced in \eqref{eq:dim}, we similarly define 
$\mathcal{Q}=\mathcal{Q}_c(1+q)$. With this substitution, the equation of state reduces to the following simplified form.
\begin{equation}
\begin{split}
&0.949 (t+1)=-0.003 \left(1q^2+2q-24.278\right) \rho ^4\\
&+0.002 \left(1q^2+2q-30.715\right) \rho ^3-0.002 \left(1q^2+2q\right) \rho ^2\\
&+0.002064 \left(1q^2+2q\right) \rho -0.00133274 \left(1q^2+2q-712.141t\right)
\end{split}
\end{equation}
This constitutes a quartic equation in the variable 
$\rho$, whose real solutions when $q\rightarrow 0$ are expressed as follows
\begin{equation}
\begin{split}
&\rho(t)=1.406 t^{2/3}-0.275 t^{4/3}-0.210 t^{5/3}+0.019 t^2\\&-2.178 \sqrt[3]{t}+0.162 t=0
\end{split}
\end{equation}
It is important to emphasize that this series expansion remains valid solely within the domain specified by the condition 
$2 q+712.141 t\neq0$.\\
In the vicinity of the critical point, the free energy takes the following form.
\begin{equation}
F=\frac{1}{\sqrt{\rho +1}}\left(\begin{aligned}
&0.467634 \rho ^4+2.05286 \rho ^3-0.554813 \rho ^2\\&+2.21925 \rho +4.4385
\end{aligned}\right)
\end{equation}
Subsequently, we evaluate the fractional derivatives of the free energy and examine their behavior in the asymptotic limit as 
$t\rightarrow 0^+$. The resulting expressions are presented below.
\[
\lim_{t \to 0^+} D_t^\alpha F(t) =
\begin{cases}
0, & \alpha < 4/3, \\
 +35.7054   , & \alpha =4/3, \\
\pm \infty, & \alpha > 4/3.
\end{cases}
\]
It is observed that the fractional derivative of the free energy of order 
$\alpha=4/3$  exhibits a discontinuity, serving as a characteristic indicator of a phase transition of the same order. By approaching the limit 
$q \rightarrow 0$, which corresponds to the critical point, we arrive at the following result.
\[
F(t) =
\begin{cases}
29.988 t^{4/3}-15.486 t^{5/3}-1.487 t^2\\-18.354t+4.438, & t>0,  \\
-41.100 t^{4/3}+52.374 t^{5/3}-71.660 t^2\\+18.354 t+4.438, & t<0.
\end{cases}
\]
\begin{figure}
\includegraphics[scale=0.5]{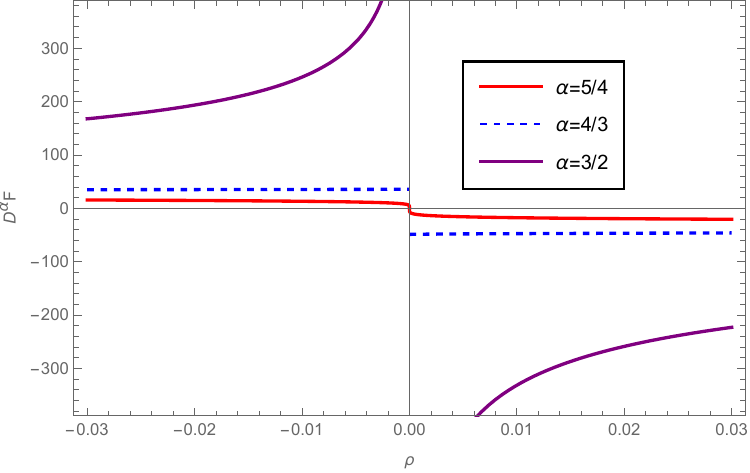}
\caption{The behavior of \( D^\alpha_t \tilde{F}(\rho) \) in the vicinity of the critical point is examined within the CFT thermodynamic framework corresponding to the RN-AdS black hole formulated with Kaniadakis statistics.
}
\label{fig:F15}
\end{figure}
Due to the emergence of 
$t^{4/3}$
  terms, a finite discontinuity manifests at 
$\alpha=4/3$
\begin{equation}
\lim_{t\rightarrow 0^-}\frac{d^\alpha \tilde{F}}{dt^\alpha}=-\frac{66.2653}{\Gamma \left(\frac{2}{3}\right)}\neq \lim_{t\rightarrow0^+}\frac{d^\alpha \tilde{F}}{dt^\alpha}=\frac{48.3494}{\Gamma \left(\frac{2}{3}\right)}
\end{equation}
In line with the generalized classification framework, the phase transition occurring at the critical point is identified as being of fractional order 
$\alpha =4/3$. The corresponding behavior is illustrated in Figure \ref{fig:F15}.\\ Additionally, it is crucial to address the special case defined by the condition 
$2 q+712.141 t=0$, which outlines a specific trajectory within the 
$(t,q)$ thermodynamic parameter space. Under this constraint, the critical point must be approached precisely along this distinct direction from both sides. Applying this condition simplifies the equation of state to the following reduced form\\
\begin{equation}
\begin{split}
&-0.002 \rho ^3 \left(126787. t^2-712.142 t-30.715\right)\\
&+0.002 \rho ^2 \left(126787. t^2-712.142 t\right)\\
&-0.002 \rho  \left(126787. t^2-712.142 t\right)\\
&+0.001 \left(126787. t^2-0.001 t\right)=0
\end{split}
\end{equation}
The expanded solutions are given as
\begin{equation}
\begin{split}
&\rho=-200.104 t^{2/3}+3.752\times 10^9 t^{4/3}+2.814\times 10^{13} t^{5/3}\\&+0.027 \sqrt[3]{t}+6.630 t\\
&\rho=200.104 t^{2/3}-3.752\times 10^9 t^{4/3}-2.814\times 10^{13} t^{5/3}\\&-0.027 \sqrt[3]{t}+6.630 t
\end{split}
\end{equation}
In this specific case, the emergence of terms with non-integer exponents results in a reduced expression for the free energy that similarly includes fractional powers. Remarkably, such features were not present in the RN-AdS black hole scenario studied before. The corresponding dimensionless form of the free energy under the constraint 
$2 q+712.141 t=0$ is provided as follows
\[
F(t) =
\begin{cases}
-0.758 t^{4/3}+5689.11 t^{5/3}-3518.15 t^2\\+0.00003 t+4.438, & t>0,  \\
0.758 t^{4/3}-5689.09 t^{5/3}+3172.02 t^2\\-0.00003 t+4.438, & t<0.
\end{cases}
\]
We compute the fractional derivatives of the free energy and analyse their behaviour in the limiting regimes as 
$t\rightarrow 0^+$. The corresponding analytical expressions are provided below
\[
\lim_{t \to 0^+} D_t^\alpha F(t) =
\begin{cases}
0, & \alpha < 4/3, \\
 +0.90311    , & \alpha =4/3, \\
\pm \infty, & \alpha > 4/3.
\end{cases}
\]
As demonstrated above and illustrated in Figure \ref{fig:F16}, the fractional derivative of the free energy at order 
$\alpha=4/3$  exhibits a finite discontinuity, signifying the emergence of a phase transition of fractional order 
$4/3$
under the specific constraint 
$2 q+712.141 t=0$. This notable behavior is absent in the corresponding RN-AdS black hole case.\\
\begin{figure}
\includegraphics[scale=0.5]{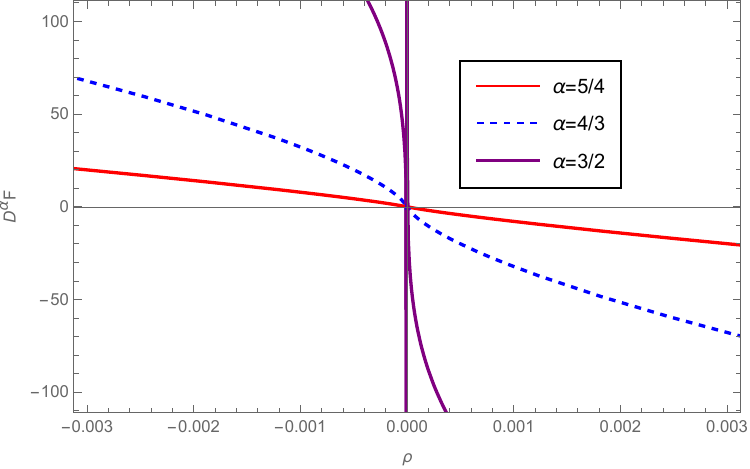}
\caption{The behavior of \( D^\alpha_t \tilde{F}(\rho) \) in the vicinity of the critical point is examined within the CFT thermodynamic framework corresponding to the RN-AdS black hole formulated with Kaniadakis statistics.
}
\label{fig:F16}
\end{figure}
We now direct our attention to the second category of Davies points, namely the second critical points using the Kaniadakis entropy, which are identified by specific thermodynamic conditions
\begin{equation}
\mathcal{S}_c=7.1223,\quad \mathcal{Q}_c=2.06541,\quad \mathcal{T}_c=0.9091
\end{equation}
Applying the definition introduced in \eqref{eq:dim}, we similarly define 
$\mathcal{Q}=\mathcal{Q}_c(1+q)$. With this reparameterization, the equation of state simplifies to the following form
\begin{equation}
\begin{split}
&0.101 \left(1q^2 +2q  +8.962 t  \right)+0.250 \rho ^4 \left(1q^2 +2q  +0.348\right)\\
&-0.222 \rho ^3 \left(1q^2 +2q  +0.189\right)+0.190 \rho ^2 \left(1q^2 +2q  \right)\\
&-0.152\rho  \left(1q^2 +2q  \right)=0
\end{split}
\end{equation}
This corresponds to a fourth-degree polynomial equation in the variable $\rho$, whose real-valued solutions are given as follows
\begin{equation}
\begin{split}
&\rho(t)=5.349 t^{2/3} +17.613 t^{4/3} +37.765 t^{5/3} +233.498 t^{7/3}\\
& +627.061 t^{8/3} +1733.15 t^3 +90.681 t^2 +2.783 \sqrt[3]{t} +9.387t 
\end{split}
\end{equation}
In the vicinity of the critical point, the free energy takes the following analytical form
\begin{equation}
\begin{split}
&F=\frac{1}{\sqrt{\rho + 1}} \left(
\begin{aligned}
&0.014 \rho^4 + 0.071 \rho^3 \\
&+ q^2 (-0.0007 \rho^2 - 0.001 \rho + 2.173)\\
&- 0.744 \rho^2 + q (-0.001 \rho^2 - 0.003 \rho + 4.347)\\
& + 2.976 \rho + 5.952
\end{aligned}
\right)
\end{split}
\end{equation}
Taking the limit 
$q\rightarrow 0$, which corresponds to the critical point, we derive the following result
\[
F(t) =
\begin{cases}
-13.5166 t^{4/3}-20.7815 t^{5/3}-30.3915 t^2\\-6.47498 t
+5.95201, & t>0,  \\
61.1474 t^{4/3}+255.912 t^{5/3}+944.761 t^2\\+6.47498 t+5.95201, & t<0.
\end{cases}
\]
Subsequently, we evaluate the fractional derivatives of the free energy and examine their behavior in the limiting case as 
$t\rightarrow 0^+$. The corresponding expressions are presented below
\[
\lim_{t \to 0^+} D_t^\alpha F(t) =
\begin{cases}
0, & \alpha < 4/3, \\
 +72.80   , & \alpha =4/3, \\
\pm \infty, & \alpha > 4/3.
\end{cases}
\]
It is observed that the fractional derivative of the free energy of order 
$\alpha=4/3$  exhibits a finite discontinuity, serving as a characteristic indicator of a phase transition of the same order.\\
Due to the occurrence of 
$t^{4/3}$  terms, a finite discontinuity emerges at the fractional order 
$\alpha=4/3$ 
\begin{equation}
\lim_{t\rightarrow 0^-}\frac{d^\alpha \tilde{F}}{dt^\alpha}=-\frac{21.7923}{\Gamma \left(\frac{2}{3}\right)}\neq \lim_{t\rightarrow0^+}\frac{d^\alpha \tilde{F}}{dt^\alpha}=\frac{98.5859}{\Gamma \left(\frac{2}{3}\right)}
\end{equation}
As per the generalized classification framework, the phase transition at the critical point is identified as being of fractional order 
$\alpha=4/3$, with its corresponding behavior illustrated in Figure \ref{fig:F17}.\\
\begin{figure}
\includegraphics[scale=0.5]{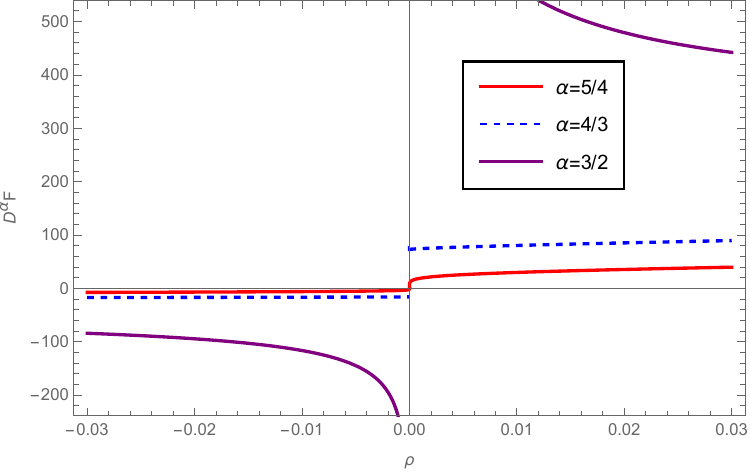}
\caption{The behavior of \( D^\alpha_t \tilde{F}(\rho) \) in the vicinity of the critical point is examined within the CFT thermodynamic framework corresponding to the RN-AdS black hole formulated with Kaniadakis statistics}
\label{fig:F17}
\end{figure}
 Additionally, it is crucial to account for the special case characterized by the constraint 
$2 q+8.962 t=0$ for the second critical point, which defines a specific trajectory within the 
$(t,q)$ parameter space. In this scenario, the critical point must be approached precisely along this unique path from both directions. Under this condition, the equation of state simplifies to the following reduced form
\begin{equation}
\begin{split}
&0.250\rho ^4 \left(20.080 t^2-8.962 t+0.348\right)\\
&-0.222 \rho ^3 \left(20.080 t^2-8.962 t+0.189\right)\\
&+0.190 \rho ^2 \left(20.080 t^2-8.962 t\right)-0.152 \rho  \left(20.080 t^2-8.962 t\right)\\
&+0.101 \left(20.080 t^2+2.534 t\right)=0
\end{split}
\end{equation}
The corresponding series-expanded solutions are delineated in the expressions provided below
\begin{equation}
\begin{split}
&\rho=592.716 t^{2/3}-2.079\times 10^{11} t^{4/3}+6.747\times 10^{15} t^{5/3}\\
&+0.0182 \sqrt[3]{t}+8.906 t\\
&\rho=-592.716 t^{2/3}+2.079\times 10^{11} t^{4/3}-6.747\times 10^{15} t^{5/3}\\&-0.0182 \sqrt[3]{t}+8.906 t
\end{split}
\end{equation}
In this specific context, the emergence of fractional power terms gives rise to a reduced free energy expression that likewise contains non-integer exponents. Significantly, such behavior was not observed in the RN-AdS black hole scenario. The dimensionless form of the free energy under the imposed constraint 
$2q+8.962t=0$ is expressed as follows
\[
F(t) =
\begin{cases}
-0.177 t^{4/3}-5788. t^{5/3}+4.027 t^2\\-1.83\times 10^{-6} t+5.952, & t>0,  \\
0.178 t^{4/3}+5788. t^{5/3}+206.728 t^2\\+1.83\times 10^{-6} t+5.952, & t<0.
\end{cases}
\]
We proceed to compute the fractional derivatives of the free energy and analyze their behavior in the limiting regime as 
$t\rightarrow 0^+$. The resulting analytical expressions are detailed below
\[
\lim_{t \to 0^+} D_t^\alpha F(t) =
\begin{cases}
0, & \alpha < 4/3, \\
 +0.212    , & \alpha =4/3, \\
\pm \infty, & \alpha > 4/3.
\end{cases}
\]
As depicted earlier and in Figure \ref{fig:F18}, the fractional derivative of the free energy of order 
$\alpha=4/3$ reveals a finite discontinuity, indicating the presence of a phase transition of the same order under the specific constraint 
$2q+8.962t=0$. This distinct behavior is notably absent in the corresponding RN-AdS black hole scenario.\\
\begin{figure}
\includegraphics[scale=0.5]{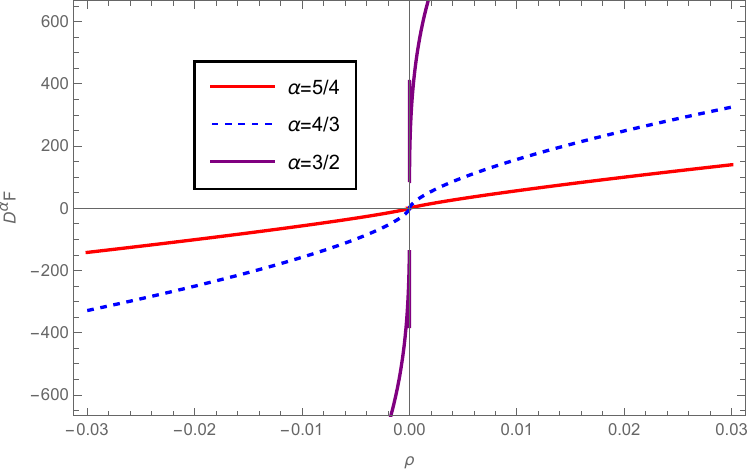}
\caption{The behavior of \( D^\alpha_t \tilde{F}(\rho) \) in the vicinity of the critical point is examined within the CFT thermodynamic framework corresponding to the RN-AdS black hole formulated with Kaniadakis statistics}
\label{fig:F18}
\end{figure}
The emergence of different fractional orders in distinct settings underscores the sensitivity of critical behavior to the underlying entropy formalism. In the case of Kaniadakis entropy, a fractional-order phase transition with 
$\alpha =4/3$ appears consistently  both generically and along the fine tuned path defined by 
$2q+8.962t=0$. This suggests that the 
$4/3$-order transition is intrinsic to the thermodynamic structure dictated by Kaniadakis statistics. In contrast, when employing the ModMax-AdS framework, the same fine-tuned direction yields a transition of order 
$\alpha =3/2$ , highlighting a qualitatively distinct critical response. Therefore, the fractional order of the phase transition is not universal but depends sensitively on the entropy model and the geometry considered.
\section{CONCLUSION}\label{sec:conclusion}
In this work, we investigate the CFT phase structure of the Reissner–Nordstr\"om–AdS (RN-AdS) black hole, ModMax-AdS black hole and RN-AdS black hole encoded with the Kaniadakis statistics  at its Davies points using the framework of the generalized Ehrenfest classification based on fractional-order derivatives. Traditionally, Davies points where the heat capacity diverges are classified into two categories: those corresponding to extrema of the temperature and those associated with inflection points (i.e., critical points). Within the scope of the conventional Ehrenfest scheme, such divergences are typically interpreted as indicators of second-order phase transitions. However, employing the fractional calculus-based generalization, we uncover that the transition occurring at the first type of Davies point is of fractional order $\alpha=3/2$ , while the second type exhibits a transition of order 
$\alpha=4/3$ . Notably, the both results is consistent with that obtained under the extended phase space formalism \cite{mengsen_ma,wang_he_ma_fractional_rnads}, thereby validating the robustness of this approach. These findings demonstrate that the generalized Ehrenfest classification provides a more refined and insightful characterization of critical phenomena in CFT thermodynamics, revealing distinctions that remain obscured in the traditional framework.\\
While it was initially believed that the equation of state alone dictates the order of a fractional phase transition, recent findings suggest a more nuanced picture. As highlighted in \cite{chabab_iraoui1}, the underlying spacetime symmetry of black holes may also play a significant role in determining the nature and order of such transitions. In our present analysis, we observe that the equations of state near both the first and second types of Davies points for all the black hole systems take the form of quartic equations. However, this observation does not, in itself, fully resolve the question of what governs the order of the phase transition. To better understand this issue, further investigations involving other black hole backgrounds and well as its CFT counterparts are necessary.\\
Although we identify a fractional-order CFT phase transition of 
$\alpha=3/2$ and $\alpha=4/3$ at the first type and second type of Davies point for all the three black hole systems, it remains unclear whether black holes with similar thermodynamic behavior exhibit the same order of transition. To date, the generalized Ehrenfest classification has predominantly been applied to phase transitions occurring specifically at Davies points. Nevertheless, its scope could be extended to provide a fresh perspective on standard first-order phase transitions as well.\\
An intriguing observation emerges when analysing the phase structure of the CFT dual to the RN-AdS black hole under a specific constraint in the thermodynamic parameter space. This constraint imposed to probe a particular trajectory toward the critical point allows us to isolate a well defined path in the multidimensional thermodynamic landscape, thereby enabling a more controlled and analytical expansion of thermodynamic quantities near criticality. Under this condition, we observe a fractional-order phase transition of order 
$\alpha=3/2$ a behaviour that is notably absent in the gravitational (bulk) formulation of the RN-AdS black hole. This suggests that certain critical features may be exclusive to the CFT thermodynamic framework, possibly due to the boundary theory's sensitivity to fluctuations or fine tuned ensemble constraints that lack a direct bulk analogue. Interestingly, a similar phenomenon is observed in the CFT analysis of the ModMax-AdS black hole, where the same $3/2$ order transition appears under analogous conditions. These results emphasize that critical behaviour in the boundary CFT can be richer or more subtle than what is apparent from the bulk perspective alone, highlighting the depth of the AdS/CFT correspondence.\\
However, the picture changes when we incorporate Kaniadakis entropy into the CFT thermodynamics of the RN-AdS black hole. The deformation introduced by this non-extensive entropy model leads to the appearance of two distinct critical points, each giving rise to a different constraint in parameter space. Unlike the previous cases, both constrained paths now yield a phase transition of order $\alpha=4/3$ consistent with the result obtained without imposing any constraint. The absence of the 
$3/2$ order transition in this setup can be attributed to the influence of the Kaniadakis parameter $\kappa$, which modifies the thermodynamic structure in such a way that the system's critical response becomes less sensitive to the path of approach to the critical point. In essence, the constraint no longer induces a bifurcation in phase behaviour because the deformation smooths out the thermodynamic curvature, preserving the dominant critical structure across paths. This contrast illustrates that the order of fractional phase transitions is not only influenced by the nature of the constraint, but also depends critically on the underlying entropy model used to describe the system.\\
 The order of a phase transition is closely tied to the critical exponents, which are experimentally measurable quantities in real physical systems. A systematic exploration of the critical exponents associated with fractional-order phase transitions in black holes may thus open new pathways for connecting black hole thermodynamics with experimentally accessible systems. These intriguing questions warrant further study in future research endeavours.

\section*{Acknowledgements} 
AB gratefully acknowledges Meng-Sen Ma for his valuable assistance and insightful guidance in this work.
%

\end{document}